\documentclass[a4paper,11pt]{article}
\usepackage{amssymb, amsmath}
\usepackage[curve]{xypic}

\providecommand{\abs}[1]{\lvert#1\rvert}

\providecommand{\Tr}{\textnormal{Tr}}

\providecommand{\Hol}{\textnormal{Hol}}

\providecommand{\pfaff}{\textnormal{pfaff}}
\providecommand{\Maps}{\textnormal{Maps}}

\catcode`\@=11

\def\section{\@startsection{section}{1}{\z@}{3.5ex plus 1ex minus
   .2ex}{2.3ex plus .2ex}{\LARGE\bf}}
\def\subsection{\@startsection{subsection}{1}{\z@}{3.5ex plus 1ex minus
   .2ex}{2.3ex plus .2ex}{\Large\bf}}
\def\subsubsection{\@startsection{subsubsection}{1}{\z@}{3.5ex plus 1ex minus
   .2ex}{2.3ex plus .2ex}{\large\bf}}

\begin{document}

\begin{titlepage}
\titlepage
\rightline{SISSA 67/2008/EP-FM}
\vskip 2.5cm
\centerline{ \bf \huge Classifying A-field and B-field configurations }
\vskip 0.7cm
\centerline{ \bf \huge in the presence of D-branes }
\vskip 1.7truecm

\begin{center}
{\bf \Large Loriano Bonora, Fabio Ferrari Ruffino and Raffaele Savelli}
\vskip 1.5cm
\em 
International School for Advanced Studies (SISSA/ISAS) \\ 
Via Beirut 2, I-34014, Trieste, Italy\\
and Istituto Nazionale di Fisica Nucleare (INFN), sezione di Trieste

\vskip 2.5cm

\large \bf Abstract
\end{center}

\normalsize We ``solve'' the Freed-Witten anomaly equation, i.e., we find a geometrical classification of the $B$-field and $A$-field configurations in the presence of D-branes that are anomaly-free. The mathematical setting being provided by the geometry of gerbes, we find that the allowed configurations are jointly described by a coset of a certain hypercohomology group. We then describe in detail various cases that arise according to such classification. As is well-known, only under suitable hypotheses the A-field turns out to be a connection on a canonical gauge bundle. 
However, even in these cases, there is a residual freedom in the choice of the bundle, naturally arising from the hypercohomological description. For a B-field which is flat on a D-brane, fractional or irrational charges of subbranes naturally appear; for a suitable gauge choice, they can be seen as arising from ``gauge bundles with not integral Chern class'': we give a precise geometric interpretation of these objects.

\vskip2cm

\vskip1.5\baselineskip

\vfill
 \hrule width 5.cm
\vskip 2.mm
{\small 
\noindent }
\begin{flushleft}
bonora@sissa.it, ferrari@sissa.it, savelli@sissa.it
\end{flushleft}
\end{titlepage}

\large

\newtheorem{Theorem}{Theorem}[section]
\newtheorem{Lemma}[Theorem]{Lemma}
\newtheorem{Corollary}[Theorem]{Corollary}
\newtheorem{Rmk}[Theorem]{Remark}
\newtheorem{Def}{Definition}[section]
\newtheorem{ThmDef}[Theorem]{Theorem - Defintion}

\tableofcontents

\newpage

\section{Introduction}

A string background characterized by a $B$-field in the presence of D-branes induces a geometry where traditional mathematical or field theoretical tools have to be updated or upgraded in order to provide an effective description. For instance, it is generally accepted that the appropriate mathematical structure underlying such backgrounds is provided by gerbes, which are generalizations of bundles. In this framework the idea of a gauge bundle associated with the worldvolume of a D-brane is not always adequate and needs to be refined. The new structure 
has to be free of Freed-Witten anomalies, which are global anomalies of the world-sheet path integral. These problems have already been analyzed in the literature, starting from the seminal paper of D.Freed and E.Witten \cite{FW}. However the analysis has been carried out in a case by case basis and a general classifying scheme is still lacking. In this paper we would like to fill in the gap and show that a mathematical tool exists that is capable of encompassing all the particular cases of backgrounds mentioned above: this is $\rm\check{C}$ech hypercohomology of sheaves.

In fact, the second hypercohomology group of some specific sheaves (characterizing the space-time and the D-brane world-volumes), which we describe below, provides us with a tool to classify ``gerbes with connection'' that are Freed-Witten anomaly free. This turns out to be the instrument we need in order to select the right ``string backgrounds with D-branes and $B$-field". To make an example which is more familiar in the physical literature, the first hypercohomology group of the same sheaves classify all the line bundles with connection, so it classifies the (classical) $U(1)$ gauge field theories. To deal with ``string backgrounds with D-branes and B-field'' we need to go one step further in hypercohomology with respect to this example.

We would like to point out that the term ``hypercohomology'' is rarely used in the physical literature, but, on the other hand, one can find interesting examples of it under a different terminology. For instance, the double BRST complex in local field theory is an example in which the famous descent equations are exactly the cocycle conditions for hypercohomology.

The second part of the paper consists in a description of the various cases which arise from this type of classification. Some of them have already been analyzed in the literature, others are new. It is, for instance, well-known that generically we cannot define a canonical gauge theory on a D-brane in the presence of a non-zero B-field, due to the freedom under large gauge transformations. Such possibility arises only if some specific conditions are satisfied. Even when this is possible the hypercohomological description tells us that there is a 
residual freedom in the choice of the bundle. 

We then describe in particular, using the hypercohomology context, the situation arising when the B-field is flat. In such cases the analysis confirms what has already been established in the literature: fractional or irrational charges of subbranes naturally appear; for a suitable gauge choice, they can be seen as arising from ``gauge bundles with 
not integral Chern class'': we give a precise geometric interpretation of these objects.

The paper is organized as follows. Section 2 is devoted to a review of the Freed-Witten anomaly. In section 3 we introduce the formalism necessary to describe holonomy for gerbes. Section 4 contains the central result of our paper: we introduce suitable hypercohomology groups and show how they classify B-field and A-field configurations in the presence of D-branes. Section 5 is devoted to a description of the possible gauge theories on D-branes in the presence of a B-field, which arise from the previous classification. In section 6 we give a geometrical description of the fractional bundles that appear in certain B-field and A-field configurations. In section 7 we briefly describe the generalization to stacks of coincident D-branes. Section 8 contains our conclusions while the two appendices are devoted to the definition of $\rm\check{C}$ech hypercohomology and gerbes with connection.

\section{The Freed-Witten anomaly}

Our aim in this paper is to classify B-field and A-field configurations in type II superstring backgrounds with a fixed set of D-branes. It is well known that to this end the appropriate mathematical framework is represented by gerbes (v. \cite{Hitchin} and \cite{Brylinski}). As line bundles on a space $X$ are characterized, up to isomorphism, by the first Chern class in $H^{2}(X, \mathbb{Z})$, gerbes are classified by the first Chern class in $H^{3}(X, \mathbb{Z})$. Analogously, as a connection on a line bundle is given by local 1-forms up to gauge transformations, a connection on a gerbe is defined by local 2-forms and 1-forms up to gauge transformations. Definitions and details used in the sequel are given in appendices \ref{AppHyperC} and \ref{AppGerbes}.

Let us consider string theory on a smooth space-time $X$ and let us consider a single smooth D-brane with world-volume $Y \subset X$. At first sight, one would expect the background to contain the following data:
\begin{itemize}
	\item on $X$ a gerbe with a connection given by the B-field, with Chern class $\zeta \in H^{3}(X, \mathbb{Z})$ and curvature $H \in \Omega^{3}(X, \mathbb{Z})$, so that $H$ is a de Rham representative of $\zeta$, i.e., $\zeta \otimes_{\mathbb{Z}} \mathbb{R} \simeq [\,H\,]_{dR}$;
	\item on $Y$ a line bundle with a connection given by the A-field.
\end{itemize}
However, as pointed out in \cite{FW}, while the assignment of the gerbe on $X$ is always given in the background, the presence of the line bundle is actually consistent only in some speficic cases, the most common being the one in which the gerbe restricted to $Y$ is geometrically trivial and $w_{2}(Y) = 0$, i.e., $Y$ is spin ($w_{2}(Y)$ is the second Stiefel-Withney class of the tangent bundle of $Y$, v. \cite{LM}). In general, there is a different object on the brane. To understand what, we start from the world-sheet path-integral.

\paragraph{}In the superstring world-sheet action there are the following terms:
\begin{equation}\label{Action}
	S \supset \biggl(\, \int d\psi \, \psi \,D_{\phi}\, \psi \,\biggr) + 2\pi \cdot \biggl(\, \int_{\Sigma} \phi^{*}B + \int_{\partial \Sigma} \phi^{*}A \,\biggr)
\end{equation}
where $\phi: \Sigma \rightarrow X$ is the embedding of the string world-sheet in the target space. The exponential of the first term is the Pfaffian of the Dirac operator coupled to $TY$ via $\phi$, thus we write:
	\[e^{iS} \supset \pfaff \, D_{\phi} \cdot \exp\biggl(\,2\pi i \cdot \int_{\Sigma} \phi^{*}B\,\biggr) \cdot \exp\biggl(\, 2\pi i \cdot \int_{\partial \Sigma} \phi^{*}A \,\biggr) \,.
\]
The Pfaffian may be problematic. In fact, evaluated in a point $\phi \in \Maps(\Sigma, X)$, it must satisfy $(\pfaff \, D_{\phi})^{2} = \det D_{\phi}$, so we have a sign ambiguity and we need a natural definition of the Pfaffian, up to an overall constant which is immaterial for the path-integral. The problem is that the Pfaffian is not a function, but it is naturally a section of a line bundle over $\Maps(\partial \Sigma, Y)$, called \emph{pfaffian line bundle}, with natural metric and \emph{flat} connection (v. \cite{Freed}). If this bundle is geometrically trivial, we can choose a flat unitary section $1$ up to an overall phase, so that we determine the Pfaffian as $\pfaff\,D_{\phi} \,/\, 1$; otherwise the latter is not well defined as a number. The first Chern class of the Pfaffian line bundle depends on $W_{3}(Y)$ (where $W_{3}(Y)$ is the integral lift of the third Stiefel-Whitney class of the tangent bundle of $Y$, i.e., the obstruction to the existence of $U(1)$-charged spinors on $Y$, v.  \cite{LM} and \cite{Hitchin}), while the holonomy depends on $w_{2}(Y)$. Thus, if the brane is spin the pfaffian is a well-defined function, otherwise the best we can do is to choose \emph{local} parallel sections so that we have a local definition of $\pfaff\,D_{\phi}$.

It turns out that the terms $\exp(\,2\pi i \cdot \int_{\Sigma} \phi^{*}B\,) \cdot \exp(\, 2\pi i \cdot \int_{\partial \Sigma} \phi^{*}A \,)$ can compensate exactly the possible ambiguity of the Pfaffian, giving rise to a well-defined path-integral, if and only if:
\begin{equation}\label{FWAnomaly}
W_{3}(Y) + \zeta\vert_{Y} = 0 \,.
\end{equation}
The class $W_{3}(Y) + \zeta\vert_{Y} \in H^{3}(Y, \mathbb{Z})$ is called \emph{Freed-Witten anomaly} (see \cite{FW}). In particular, $\zeta\vert_{Y}$ must be a torsion class since $W_{3}(Y)$ is, so that $[\,H\vert_{Y}\,]_{dR} = 0$.

Taking this picture into account, we now describe the geometrical meaning of the terms $\exp(\,2\pi i \cdot \int_{\Sigma} \phi^{*}B\,) \cdot \exp(\, 2\pi i \cdot \int_{\partial \Sigma} \phi^{*}A \,)$, from which the classifying group of the $B$-field and $A$-field configurations will naturally arise.

\section{Holonomy and Wilson loop}

The purpose of this preliminary section is to give a precise definition of the holonomy integrals that appear in \eqref{Action}. Given the complexity of the definition for gerbes, we start with a description of the more familiar subject of holonomy for line bundles.

\subsection{Line bundles}

\subsubsection{Global description}

Let us consider a line bundle with connection $(L, \nabla)$ on $X$ and a \emph{closed} curve $\gamma: S^{1} \rightarrow X$ with a fixed point $x = \gamma(e^{2\pi i \cdot t})$: parallel transport along $\gamma$ gives a linear map $t_{x}: L_{x} \rightarrow L_{x}$, which can be thought of as a number $\Hol_{\nabla}(\gamma) \in S^{1}$ thanks to the canonical isomorphism $L_{x}^{\checkmark} \otimes L_{x} \simeq \mathbb{C}$ given by $\varphi \otimes v \simeq \varphi(v)$ (such a number is independent of the chosen point $x$). Thus, denoting by $LX$ the loop space of $X$, parallel transport defines a function $\Hol_{\nabla}: LX \rightarrow S^{1}$ called \emph{holonomy} of $\nabla$.

What can we say about open curves? Given a curve $\gamma: [\,0,1] \rightarrow X$, let us put $x = \gamma(0)$ and $y = \gamma(1)$: parallel transport defines a linear map $t_{x,y}: L_{x} \rightarrow L_{y}$, which is no longer canonically a number, since $L_{x}^{\checkmark} \otimes L_{y}$ is not canonically isomorphic to $\mathbb{C}$. Thus, given a curve $\gamma \in CX$, $CX$ being the space of open curves on $X$, holonomy is an element of a 1-dimensional vector space $CL_{\gamma} = L_{x}^{\checkmark} \otimes L_{y}$: we now describe this vector space as the fiber over $\gamma$ of a line bundle $CL \rightarrow CX$, so that holonomy defines a section of $CL$. In fact, let us consider the bundle $L^{\checkmark} \boxtimes L \rightarrow X \times X$, i.e., $L^{\checkmark} \boxtimes L = \pi_{1}^{*}L^{\checkmark} \otimes \pi_{2}^{*}L$ for $\pi_{1}, \pi_{2}: X \times X \rightarrow X$ the projections to the first and second factor, respectively. We have a natural map $\pi: CX \rightarrow X \times X$ given by $\pi(\gamma) = (\gamma(0), \gamma(1))$, so that we can define $CL = \pi^{*}(L^{\checkmark} \boxtimes L)$. By construction $CL_{\gamma} = (L^{\checkmark} \boxtimes L)_{\pi(\gamma)} = (L^{\checkmark} \boxtimes L)_{(\gamma(0),\gamma(1))} = L^{\checkmark}_{\gamma(0)} \otimes L_{\gamma(1)}$, so we obtain exactly the desired fiber. Thus holonomy defines a section $\Hol_{\nabla}: CL \rightarrow CX$. Moreover $c_{1}(CL) = \pi^{*}(\pi_{2}^{*}\,c_{1}(L) - \pi_{1}^{*}\,c_{1}(L))$.

As one can see from the expression of $c_{1}(CL)$, if $L$ is trivial so is $CL$. There is more: \emph{a trivialization of $L$ determines a trivialization of $\,CL$}. In fact, if $s: X \rightarrow L$ is a global section, it determines canonically a global section $s^{\checkmark}: X \rightarrow L^{\checkmark}$ given by $s^{\checkmark}(s) = X \times \{1\}$, thus a section $s^{\checkmark} \boxtimes s: X \times X \rightarrow L^{\checkmark} \boxtimes L$, thus, by pull-back, a global section $\pi^{*}(s^{\checkmark} \boxtimes s): CX \rightarrow CL$. What is happening geometrically? A global section $s: X \rightarrow L$ provides a way to identify the fibers of $L$, hence a linear map $L_{x} \rightarrow L_{y}$ becomes the number $\lambda$ such that $s_{x} \rightarrow \lambda \cdot s_{y}$. Thus, \emph{for a trivial bundle with a fixed global section, holonomy is a well-defined function also over the space of open curves}.

\paragraph{}Similarly, \emph{a system of local sections of $L$, with respect to a good cover $\mathfrak{U} = \{U_{i}\}_{i \in I}$, determines a system of local sections of $CL$}, with respect to the cover $\mathfrak{V}$ defined in the following way:
\begin{itemize}
	\item let us fix a triangulation $\tau$ of $S^{1}$, i.e., a set of vertices $\sigma^{0}_{1}, \ldots, \sigma^{0}_{l} \in S^{1}$ and of edges $\sigma^{1}_{1}, \ldots, \sigma^{1}_{l} \subset S^{1}$ such that $\partial \sigma^{1}_{i} = \sigma^{0}_{i+1} - \sigma^{0}_{i}$ for $1 \leq i < l$ and $\partial \sigma^{1}_{l} = \sigma^{0}_{1} - \sigma^{0}_{l}$;
	\item we consider the following set of indices:
	\[J = \left\{(\tau, \varphi): \quad \begin{array}{l}
	    \bullet \textnormal{ $\tau = \{\sigma^{0}_{1}, \ldots, \sigma^{0}_{l(\tau)}; \sigma^{1}_{1}, \ldots, \sigma^{1}_{l(\tau)}\}$ is a triangulation of $S^{1}$} \\
	    \bullet \textnormal{ $\varphi: \{1, \ldots, l(\tau)\} \longrightarrow I$ is a function}
	\end{array} \right\} \; ;
\]
	\item we obtain the covering $\mathfrak{V} = \{V_{(\tau,\sigma)}\}_{(\tau,\sigma) \in J}$ of $LX$ by:
	\[V_{(\tau, \varphi)} = \{\gamma \in LX: \; \gamma(\sigma^{1}_{i}) \subset U_{\varphi(i)} \} \; .
\]
\end{itemize}
Let us consider $\gamma \in V_{(\tau, \varphi)}$: then $L_{\gamma(0)}^{\checkmark} \otimes L_{\gamma(1)}$ is isomorphic to $\mathbb{C}$ via $s_{\varphi(1)}$ and $s_{\varphi(l(\tau))}$, so that we have a local trivialization $V_{(\tau, \varphi)} \times \mathbb{C}$, giving a local section $V_{(\tau, \varphi)} \times \{1\}$. Thus, we can describe transition functions of $CL$ for $\mathfrak{V}$ in terms of the ones of $L$ for $\mathfrak{U}\,$. In particular, the local expression of parallel transport along $\gamma$ with respect to the fixed local sections is given by $\{\rho_{(\tau,\varphi)}\}$ such that $t_{\gamma(0), \gamma(1)}(x, z)_{\varphi(1)} = (x, \rho_{(\tau,\varphi)} \cdot z)_{\varphi(l)}$. Then, if $\gamma \in V_{(\tau, \varphi)} \cap V_{(\tau', \varphi')}$, we have, with respect to the second chart, $t_{\gamma(0), \gamma(1)}(x, z)_{\varphi'(1)} = (x, \rho_{(\tau',\varphi')} \cdot z)_{\varphi'(l')}$. Then, since $(x, z)_{\varphi(1)} = (x, g_{\varphi(1), \varphi'(1)} \cdot z)_{\varphi'(1)}$, one has:
	\[\begin{split}
	&t_{\gamma(0), \gamma(1)}(x, z)_{\varphi(1)} = (x, \rho_{(\tau,\varphi)} \cdot z)_{\varphi(l)} = (x, g_{\varphi(l), \varphi'(l')} \cdot \rho_{(\tau,\varphi)} \cdot z)_{\varphi'(l')}\\
	&t_{\gamma(0), \gamma(1)}(x, g_{\varphi(1), \varphi'(1)} \cdot z)_{\varphi'(1)} = (x, \rho_{(\tau',\varphi')} \cdot g_{\varphi(1), \varphi'(1)} \cdot z)_{\varphi'(l')}
\end{split}\]
so that $g_{\varphi(l), \varphi'(l')} \cdot \rho_{(\tau,\varphi)} = \rho_{(\tau',\varphi')} \cdot g_{\varphi(1), \varphi'(1)}$, thus, $\rho_{(\tau,\varphi)} = \rho_{(\tau',\varphi')} \cdot (g_{\varphi(l), \varphi'(l')}^{-1} \cdot g_{\varphi(1), \varphi'(1)})$. Hence the transition functions of $CL$ are exactly $g_{(\tau, \varphi), (\tau', \varphi')}(\gamma) := g_{\varphi(l), \varphi'(l')}^{-1}\gamma(1) \cdot g_{\varphi(1), \varphi'(1)}\gamma(0)$. In particular, a trivialization $g_{ij} = g_{i}^{-1}g_{j}$ of $L$ determines a trivialization of $CL$ given by $g_{(\tau, \varphi), (\tau', \varphi')} = g_{(\tau, \varphi)}^{-1}g_{(\tau', \varphi')}$ for $g_{(\tau, \varphi)}(\gamma) = g_{\varphi(1)}\gamma(0) \cdot g_{\varphi(l)}\gamma(1)^{-1}$, as it is easy to verify.

\paragraph{}We can generalize a little bit this construction: let us consider a line bundle $L \rightarrow X$ and a subset $Y \subset X$: we can consider the space $C_{Y}X$ of open curves in $X$ with boundary in $Y$, i.e. such that $\gamma(0), \gamma(1) \in Y$. In this case, we have $\pi: C_{Y}X \rightarrow Y \times Y$ and holonomy is a section of the bundle $C_{Y}L = \pi^{*}({L\vert_{Y}}^{\checkmark} \boxtimes L\vert_{Y})$. Thus, to have a function we only need the triviality of $L\vert_{Y}$ and a global section of its, it is not necessary that the whole $L$ is trivial; similarly, to have a set of local sections of $C_{Y}L$ we just need a set of local sections of $L\vert_{Y}$.

\subsubsection{Local description}

We can now express the holonomy using local expression of the connection, so that we can generalize it to gerbes. Considering the covering $\mathfrak{V}$ of $LX$ previously defined, for a closed curve $\gamma \in V_{(\tau, \varphi)}$ we define\footnote{We consider the index $i$ of the triangulation as a cyclic index, thus $l+1 = 1$.}:
\begin{equation}\label{WilsonLoop}
\int_{\gamma} A \;:=\; \sum_{i=1}^{l(\tau)} \;\; \biggl[ \, \biggl( \int_{\gamma(\sigma^{1}_{i})}A_{\varphi(i)} \biggr) + \textstyle\frac{1}{2\pi i}\displaystyle \log g_{\varphi(i),\varphi(i+1)}\bigl(\gamma(\sigma^{0}_{i+1})\bigr) \, \biggr]
\end{equation}
and one can prove that this is a well-defined function in $\mathbb{R} / \mathbb{Z}$. Let us stress that the definition of the holonomy depends not only on the local connetion $\{A_{\alpha}\}$ but also on the cocycle $\{g_{\alpha\beta}\}$.

\paragraph{}For $\gamma$ open we must skip the last transition function. First of all we describe an analogous open cover for the space of open curves $CX$:
\begin{itemize}
	\item let us fix a triangulation $\tau$ of $[\,0,1]$, i.e., a set of vertices $\sigma^{0}_{1}, \ldots, \sigma^{0}_{l}, \sigma^{0}_{l+1} \in [\,0,1]$ and of edges $\sigma^{1}_{1}, \ldots, \sigma^{1}_{l} \subset [\,0,1]$ such that:
\begin{itemize}
	\item $\partial \sigma^{1}_{i} = \sigma^{0}_{i+1} - \sigma^{0}_{i}$ for $1 \leq i \leq l$;
	\item $\sigma^{0}_{1} = 0$ and $\sigma^{0}_{l+1} = 1$; these are called \emph{boundary vertices};
\end{itemize}
	\item we consider the following set of indices:
	\[J = \left\{(\tau, \varphi): \; \begin{array}{l}
	    \bullet \textnormal{ $\tau = \{\sigma^{0}_{1}, \ldots, \sigma^{0}_{l(\tau)}, \sigma^{0}_{l(\tau) + 1}; \sigma^{1}_{1}, \ldots, \sigma^{1}_{l(\tau)}\}$ is a triangulation of $[\,0,1]$} \\
	    \bullet \textnormal{ $\varphi: \{1, \ldots, l(\tau)\} \longrightarrow I$ is a function}
	\end{array} \right\} \; ;
\]
	\item we obtain a covering $\{V_{(\tau,\sigma)}\}_{(\tau,\sigma) \in J}$ of $CX$ by:
	\[V_{(\tau, \varphi)} = \{\gamma \in CX: \; \gamma(\sigma^{1}_{i}) \subset U_{\varphi(i)} \} \; .
\]
\end{itemize}
Thus, we define:
\begin{equation}\label{WilsonLineBundle}
\int_{\gamma} A \;:=\; \Biggl(\sum_{i=1}^{l(\tau)-1} \; \int_{\gamma(\sigma^{1}_{i})}A_{\varphi(i)} + \log g_{\varphi(i),\varphi(i+1)} \bigl(\gamma(\sigma^{0}_{i+1})\bigr) \Biggr) + \int_{\gamma(\sigma^{1}_{l})}A_{\varphi(l)} \; .
\end{equation}
In this case the integral is not well-defined as a function, but, as we have seen, it is \emph{a section of a line bundle} $CL \rightarrow CX$ with transition functions $\tilde{g}_{(\tau, \varphi), (\tau', \varphi')}(\gamma) = g_{\varphi(l),\varphi'(l)}\gamma(1)^{-1} \cdot g_{\varphi(1),\varphi'(1)}\gamma(0)$. If, for a submanifold $Y \subset X$, we ask that $\partial \gamma \subset Y$ and we choose a trivialization of $L\vert_{Y}$ given by $g_{\alpha\beta}(y) = g_{\alpha}^{-1}(y)\cdot g_{\beta}(y)$, we can express the transition functions of $CL$ as $\tilde{g}_{(\tau, \varphi), (\tau', \varphi')}(\gamma) = (g_{\varphi(l)}\gamma(1) \cdot g_{\varphi(1)}\gamma(0)^{-1}) \cdot (g_{\varphi'(l)}\gamma(1) \cdot g_{\varphi'(1)}\gamma(0)^{-1})^{-1}$, thus we obtain a trivialization of $CL$ given by  $\tilde{g}_{(\tau, \varphi)}(\gamma) = g_{\varphi(l)}\gamma(1)^{-1} \cdot g_{\varphi(1)}\gamma(0)$. With respect to this trivialization, holonomy becomes a function given by:
\begin{equation}\label{WilsonLine}
\begin{split}
	\int_{\gamma} A \;:=\; \Biggl(\sum_{i=1}^{l(\tau)-1} \; \int_{\gamma(\sigma^{1}_{i})} A_{\varphi(i)} + \log g_{\varphi(i),\varphi(i+1)}&\bigl(\gamma(\sigma^{0}_{i+1})\bigr) \Biggr) + \int_{\gamma(\sigma^{1}_{l})}A_{\varphi(l)}\\
	&+ \,\textstyle \frac{1}{2\pi i} \displaystyle\, \bigl( \, \log g_{\varphi(l)}(\gamma(1)) - \log g_{\varphi(1)}(\gamma(0)) \, \bigr) \; .
\end{split}
\end{equation}

\subsubsection{Cohomology classes and cocycles}\label{CocyclesCohomology}

We remark the following facts, which will be useful later to better figure out by analogy the case of gerbes. Let us fix a good cover $\mathfrak{U} = \{U_{\alpha}\}_{\alpha \in I}$ on a space $X$:
\begin{itemize}
	\item when we specify a \emph{cohomology class} $\alpha = [\,\{g_{\alpha\beta}\}\,] \in \check{H}^{1}(\mathfrak{U}, C^{\infty}(\,\cdot\,, \mathbb{C}^{*}))$, we associate to it an \emph{equivalence class up to isomorphism} of line bundles, represented by\footnote{This equivalence class is much larger than the class made by the bundles of the form \eqref{BundleCharts} for the various representatives $\{g_{\alpha\beta}\}$ of $\alpha$, since there are all the bundles which are not of the form \eqref{BundleCharts} but only isomorphic to one of them.}:
	\begin{equation}\label{BundleCharts}
	\Bigl( \, \bigsqcup \, (\,U_{\alpha} \times \mathbb{C}\,) \, \Bigr) \Big/ \sim \;, \qquad (x, z)_{\alpha} \sim (x, g_{\alpha\beta}(x) \cdot z)_{\beta}, \textnormal{ for }x \in U_{\alpha\beta} \; ;
\end{equation}
	\item when we specify a \emph{cocycle} $\{g_{\alpha\beta}\} \in \check{Z}^{1}(\mathfrak{U}, C^{\infty}(\,\cdot\,, \mathbb{C}^{*}))$, we associate to it the \emph{equivalence class} of a line bundle \emph{with a fixed set of local sections} $\{s_{\alpha}: X \rightarrow L\}$ \emph{up to isomorphism with relative pull-back of the sections}, such that $g_{\alpha\beta} = s_{\alpha}/s_{\beta}$. In this case we have dependence on the covering $\mathfrak{U}$, but this is obvious since the local sections themselves determines the covering by their domains. We have a canonical representative for each of these classes given by \eqref{BundleCharts}.
\end{itemize}

If we give a line bundle $L$ with a fixed set of local sections $\{s_{\alpha}: U_{\alpha} \rightarrow L\}$, it is canonically isomorphic to a line bundle of the form \eqref{BundleCharts} for $g_{\alpha\beta} = s_{\alpha} / s_{\beta}$ (of course the sections $\{s_{\alpha}\}$ do not make $\{g_{\alpha\beta}\}$ a coboundary since they are not functions, they are sections of a bundle). The isomorphism is simply given by $\varphi(s_{\alpha})_{x} = (x, 1)_{\alpha}$, and it can be applied to any bundle isomorphic to $L$ with the pull-back of the sections $\{s_{\alpha}\}$.

\subsection{Gerbes}

The situation of gerbes is analogous to the one of bundles. In particular, the holonomy of a gerbe over a closed surface is a well defined function, while the holonomy for a surface with boundary $\Sigma$ is a section of a bundle over the space $\Maps(\Sigma, X)$. If we consider the maps such that $\phi(\partial\Sigma) \subset Y$, then a trivialization of the gerbe on $Y$, if it exists, determines a trivialization of the bundle, so that holonomy becomes a well-defined function.

\subsubsection{Closed surfaces}

\begin{Def} Given a topological space $X$ and a closed compact surface $\Sigma$, the \emph{space of maps from $\Sigma$ to $X$}, called $\Sigma X$, is the set of continuous maps:
	\[\Gamma: \Sigma \longrightarrow X
\]
equipped with the compact-open topology.
\end{Def}
We now describe a natural open covering for the space of maps. In particular:
\begin{itemize}
	\item let us fix a triangulation $\tau$ of $\Sigma$, i.e.:
\begin{itemize}
	\item a set of vertices $\sigma^{0}_{1}, \ldots, \sigma^{0}_{l} \in \Sigma$;
	\item a subset $E \subset \{1, \ldots, l\}^{2}$, determining a set of oriented edges $\{\sigma^{1}_{(a,b)} \subset \Sigma\}_{(a,b) \in E}$ such that $\partial \sigma^{1}_{(a,b)} = \sigma^{0}_{b} - \sigma^{0}_{a}$; if $(a,b) \in E$ then $(b,a) \notin E$ and we declare $\sigma^{1}_{(b,a)} := -\sigma^{1}_{(a,b)}$;
	\item a subset $T \subset \{1, \ldots, l\}^{3}$, determining a set of oriented triangles $\{\sigma^{2}_{(a,b,c)} \subset \Sigma\}_{(a,b,c) \in T}$ such that $\partial \sigma^{2}_{(a,b,c)} = \sigma^{1}_{(a,b)} + \sigma^{1}_{(b,c)} + \sigma^{1}_{(c,a)}$; given $a,b,c$ only one permutation of them belongs to $T$ and for a permutation $\rho$ we declare $\sigma^{2}_{\rho(a),\rho(b),\rho(c)} := (-1)^{\rho}\sigma^{2}_{(a,b,c)}$;
\end{itemize}
satisfying the following conditions:
\begin{itemize}
	\item every point $P \in \Sigma$ belongs to at least one triangle, and if it belongs to more than one triangle then it belongs to the boundary of each of them;
	\item every edge $\sigma^{1}_{(a,b)}$ lies in the boundary of exactly two triangles $\sigma^{2}_{(a,b,c)}$ and $\sigma^{2}_{(b, a, d)}$, inducing on it opposite orientations, and $\sigma^{2}_{(a,b,c)} \cap \sigma^{2}_{(b, a, d)} = \sigma^{1}_{(a,b)}$; if a point $p \in \Sigma$ belongs to an edge $\sigma^{1}_{(a,b)}$ and it's not a vertex, than the only two triangles containing it are the ones having $\sigma^{1}_{(a,b)}$ as common boundary; thus, there exists a function $b: E \rightarrow T^{2}$ such that $\sigma^{1}_{(a,b)} \subset \partial \sigma^{2}_{b^{1}(a,b)}$ and $-\sigma^{1}_{(a,b)} \subset \partial \sigma^{2}_{b^{2}(a,b)}$;
	\item for every vertex $\sigma^{0}_{i}$ there exists a finite set of triangles $\{\sigma^{2}_{(i,a_{1},a_{2})}, \ldots, \sigma^{2}_{(i,a_{k_{i}},a_{1})}\}$ having $\sigma^{0}_{i}$ as vertex, such that $\sigma^{2}_{(i,a_{j},a_{j+1})} \cap \sigma^{2}_{(i,a_{j+1},a_{j+2})} = \sigma^{1}_{(i,a_{j+1})}$ (we use the notation $k_{i} + 1 = 1$), these triangles are the only one containing $\sigma^{0}_{i}$ and their union is a neighborhood of it; thus, there exists a function $B: \{1, \ldots, l\} \rightarrow \coprod_{i=1}^{l} T^{k_{i}}$, such that $B(i) \in T^{k_{i}}$ and $B(i) = \{\sigma^{2}_{(i,a_{1},a_{2})}, \ldots, \sigma^{2}_{(i,a_{k_{i}},a_{1})}\}$;
\end{itemize}
	\item we consider the following set of indices:
	\[J = \left\{(\tau, \varphi): \quad \begin{array}{l}
	    \bullet \textnormal{ $\tau = \bigl\{\sigma^{0}_{1}, \ldots, \sigma^{0}_{l(\tau)}, E, T \bigr\}$ is a triangulation of $\Sigma$} \\
	    \bullet \textnormal{ $\varphi: T \longrightarrow I$ is a function}
	\end{array} \right\}		 
\]
and a covering $\{V_{(\tau,\sigma)}\}_{(\tau,\sigma) \in J}$ of $\Sigma X$ is given by:
	\[V_{(\tau, \varphi)} = \{\Gamma \in \Sigma X: \; \Gamma(\sigma^{2}_{(a,b,c)}) \subset U_{\varphi(a,b,c)} \}.
\]
\end{itemize}
One can prove that these sets are open in the compact-open topology and that they cover $\Sigma X$.

For a fixed $\Gamma \in \Sigma X$, there exists $(\tau, \varphi) \in J$ such that $\Gamma \in V_{(\tau, \varphi)}$. The function $\varphi: T \rightarrow I$ induces two functions:
\begin{itemize}
	\item $\varphi^{E}: E \rightarrow I^{2}$, given by $\varphi^{E}(a,b) = \bigl( \varphi(b^{1}(a,b)), \varphi(b^{2}(a,b) \bigr)$;
	\item $\varphi^{V}: \{1, \ldots, l\} \rightarrow \coprod_{i=1}^{l} (I^{3})^{k_{i}-2}$, such that $\varphi^{V}(i) \in (I^{3})^{k_{i}-2}$ and $\bigl(\varphi^{V}(i)\bigr)^{j} = \bigl( \varphi(B^{1}(i)),$ $\varphi(B^{j}(i)), \varphi(B^{j+1}(i)) \bigr)$.
\end{itemize}
We define:
\begin{equation}\label{WilsonLoopGerbe}
\begin{split}
\int_{\Gamma} B \;:=\; \sum_{(a,b,c) \in T_{\tau}} \; \int_{\Gamma(\sigma^{2}_{(a,b,c)})} B_{\varphi(a,b,c)} \; + \; \sum_{(a,b) \in E_{\tau}} \;& \int_{\Gamma(\sigma^{1}_{(a,b)})} \Lambda_{\varphi^{E}(a,b)}\\
& + \; \sum_{i=1}^{l} \sum_{j=1}^{k_{i}} \; \log g_{(\varphi^{V}(i))^{j}}\bigl(\Gamma(\sigma^{0}_{i})\bigr) \; .
\end{split}
\end{equation}
The last term needs some clarifications: we briefly discuss it. The logarithm can be taken since we have chosen a good covering, so the intersections are contractible. Of course, it's  defined up to $2\pi i \,\mathbb{Z}$, so the quantity that can be well-defined as a number is $\exp\bigl(\int_{\Gamma} B\bigr)$. The sum is taken in the following way: we consider the star of triangles having $\sigma^{0}_{i}$ as common vertex (each of them associated to a chart via $\varphi$) and, since we are considering 0-simplices, that corresponds to 2-cochains, we consider the possible triads with first triangle fixed $\bigl(\varphi^{V}(i)\bigr)^{j} = \bigl( \varphi(B^{1}(i)),$ $\varphi(B^{j}(i)), \varphi(B^{j+1}(i)) \bigr)$ and sum over them. The fact we fixed $B^{1}(i)$ as first triangle has no effect, since we could consider any other possibility $\bigl(\varphi^{V}_{\alpha}(i)\bigr)^{j} = \bigl( \varphi(B^{\alpha}(i)),$ $\varphi(B^{j}(i)), \varphi(B^{j+1}(i)) \bigr)$. In fact, by cocycle condition with indices $(1, i, i+1, \alpha)$ we have that $g_{1,i+1,\alpha} \cdot g_{1, i, i+1} = g_{i,i+1,\alpha} \cdot g_{1, i, \alpha}$, thus $g_{\alpha,i,i+1} = g_{1, i, \alpha}^{-1} \cdot g_{1, i, i+1} \cdot g_{1,i+1,\alpha}$, but in the cyclic sum the extern terms simplify, hence the sum involving $g_{\alpha,i,i+1}$ is equal to the sum involving $g_{1,i,i+1}$. Finally, we sumed over $j = 1, \ldots, k_{i}$, but for $j = 1$ and $j = k_{i}$ we obtain trivial terms, hence the real sum is for $j = 2, \ldots, k_{i}-1$.

\subsubsection{Surfaces with boundary}

\begin{Def} Given a topological space $X$ and a compact surface \emph{with boundary} $\Sigma$, the \emph{space of maps from $\Sigma$ to $X$}, called $\Sigma X$, is the set of continuous maps:
	\[\Gamma: \Sigma \longrightarrow X
\]
equipped with the compact-open topology.
\end{Def}
As before:
\begin{itemize}
	\item let us fix a triangulation $\tau$ of $\Sigma$, i.e.:
	\begin{itemize}
	\item a set of vertices $\sigma^{0}_{1}, \ldots, \sigma^{0}_{l} \in \Sigma$;
	\item a subset $E \subset \{1, \ldots, l\}^{2}$, determining a set of oriented edges $\{\sigma^{1}_{(a,b)} \subset \Sigma\}_{(a,b) \in E}$ such that $\partial \sigma^{1}_{(a,b)} = \sigma^{0}_{b} - \sigma^{0}_{a}$; if $(a,b) \in E$ then $(b,a) \notin E$ and we declare $\sigma^{1}_{(b,a)} := -\sigma^{1}_{(a,b)}$;
	\item a subset $T \subset \{1, \ldots, l\}^{3}$, determining a set of oriented triangles $\{\sigma^{2}_{(a,b,c)} \subset \Sigma\}_{(a,b,c) \in T}$ such that $\partial \sigma^{2}_{(a,b,c)} = \sigma^{1}_{(a,b)} + \sigma^{1}_{(b,c)} + \sigma^{1}_{(c,a)}$; given $a,b,c$ only one permutation of them belongs to $T$ and for a permutation $\rho$ we declare $\sigma^{2}_{\rho(a),\rho(b),\rho(c)} := (-1)^{\rho}\sigma^{2}_{(a,b,c)}$;
\end{itemize}
satisfying the usual conditions for triangulations; there exists a partition $E = BE \,\dot{\cup}\, IE$ in \emph{boundary edges} and \emph{internal edges}, and two functions:
\begin{itemize}
	\item $b: IE \rightarrow T^{2}$ such that $\sigma^{1}_{(a,b)} \subset \partial \sigma^{2}_{b^{1}(a,b)}$ and $-\sigma^{1}_{(a,b)} \subset \partial \sigma^{2}_{b^{2}(a,b)}$;
	\item $b: BE \rightarrow T$ such that $\sigma^{1}_{(a,b)} \subset \partial \sigma^{2}_{b(a,b)}$;
\end{itemize}
moreover, there exists a partition $\{0, \ldots, l\} = BV \,\dot{\cup}\, IV$ in \emph{boundary vertices} and \emph{internal vertices}, such that:
\begin{itemize}
	\item for $i \in IV$, there exists a finite set of triangles $\{\sigma^{2}_{(i,a_{1},a_{2})}, \ldots, \sigma^{2}_{(i,a_{k_{i}},a_{1})}\}$ having $\sigma^{0}_{i}$ as vertex; $\sigma^{2}_{(i,a_{j},a_{j+1})} \cap \sigma^{2}_{(i,a_{j+1},a_{j+2})} = \sigma^{1}_{(i,a_{j+1})}$ with a cyclic order (i.e., we use the notation $k_{i} + 1 = 1$); these triangles are the only ones containing $\sigma^{0}_{i}$ and their union is a neighborhood of it; thus, there exists a function $B: IV \rightarrow \coprod_{i\in IV} T^{k_{i}}$, such that $B(i) \in T^{k_{i}}$ and $B(i) = \{\sigma^{2}_{(i,a_{1},a_{2})}, \ldots, \sigma^{2}_{(i,a_{k_{i}},a_{1})}\}$.
	\item for $i \in BV$, there exists a finite set of triangles $\{\sigma^{2}_{(i,a_{1},a_{2})}, \ldots, \sigma^{2}_{(i,a_{k_{i}-1},a_{k_{i}})}\}$ (without $\sigma^{2}_{(i,a_{k_{i}},a_{1})}$) having $\sigma^{0}_{i}$ as vertex; $\sigma^{2}_{(i,a_{j},a_{j+1})} \cap \sigma^{2}_{(i,a_{j+1},a_{j+2})} = \sigma^{1}_{(i,a_{j+1})}$ for $1 < i < k_{i}$, these triangles are the only ones containing $\sigma^{0}_{i}$ and their union is a neighborhood of it; thus, there exists a function $B: BV \rightarrow \coprod_{i=\in BV} T^{k_{i}-1}$, such that $B(i) \in T^{k_{i}-1}$ and $B(i) = \{\sigma^{2}_{(i,a_{1},a_{2})}, \ldots, \sigma^{2}_{(i,a_{k_{i}-1},a_{k_{i}})}\}$;
\end{itemize}

	\item we consider the following set of indices:
	\[J = \left\{(\tau, \varphi): \quad \begin{array}{l}
	    \bullet \textnormal{ $\tau = \bigl\{\sigma^{0}_{1}, \ldots, \sigma^{0}_{l(\tau)}, E, T \bigr\}$ is a triangulation of $\Sigma$} \\
	    \bullet \textnormal{ $\varphi: T \longrightarrow I$ is a function}
	\end{array} \right\}		 
\]
and a covering $\{V_{(\tau,\sigma)}\}_{(\tau,\sigma) \in J}$ of $\Sigma X$ is given by:
	\[V_{(\tau, \varphi)} = \{\Gamma \in \Sigma X: \; \Gamma(\sigma^{2}_{(a,b,c)}) \subset U_{\varphi(a,b,c)} \}.
\]
\end{itemize}
One can prove that these sets are open in the compact-open topology and that they cover $\Sigma X$.

\paragraph{}For the holonomy in this case, the only possibility is to use the same definition as for closed surfaces, omitting the boundary edges and vertices in the integration. This forbids the well-definedness of the integral as a function. We obtain a \emph{line bundle} $\tilde{L}$ over the space of maps $\Maps(\partial\Sigma, Y)$ with the following properties (we call $\mathcal{G}$ the gerbe):
\begin{itemize}
	\item $c_{1}(\tilde{L})$ depends on $c_{1}(\mathcal{G})$, thus, if $\mathcal{G}$ is trivial then $\tilde{L}$ is trivial too;
	\item a particular realization of $\mathcal{G}$ as a $\rm\check{C}$ech hypercocycle (see appendices \ref{AppHyperC} and \ref{AppGerbes} for notations) determines a realization of $\tilde{L}$ as $\rm\check{C}$ech cocycle; in particular, if $\mathcal{G}$ is of the form $\{g_{\alpha\beta\gamma}, 0, B\}$ with $g_{\alpha\beta\gamma}$ constant, we obtain a realization of $\tilde{L}$ with constant transition functions whose class in $H^{2}(\Maps(\partial\Sigma, Y), S^{1})$ depends on $[\,\{g_{\alpha\beta}\}\,] \in H^{2}(Y, S^{1})$. In particular, for a realization of the form $\{\eta_{\alpha\beta\gamma}, 0, B\}$ with $[\,\eta_{\alpha\beta\gamma}\,] = w_{2}(Y)$, we obtain a realization of the $\tilde{L}$ with the same class as the parallel sections of the Pfaffian line bundle.
\end{itemize}

\paragraph{}One can prove that the function for a specific trivialization can be obtained in the following way. Let us consider a trivial gerbe $\{g_{\alpha\beta\gamma}\} \in \check{B}^{2}(X, \underline{S}^{1})$, and let $g_{\alpha\beta\gamma} = g_{\alpha\beta} \cdot g_{\beta\gamma} \cdot g_{\gamma\alpha}$. We have:
	\[\begin{split}
	&B_{\alpha} - B_{\beta} = d\Lambda_{\alpha\beta}\\
	&\Lambda_{\alpha\beta} +  \Lambda_{\beta\gamma} + \Lambda_{\gamma\alpha} = d \log g_{\alpha\beta} + d \log g_{\beta\gamma} + d \log g_{\gamma\alpha}\\
	&\bigl( \Lambda_{\alpha\beta} - d \log g_{\alpha\beta} \bigr) + \bigl( \Lambda_{\beta\gamma} - d \log g_{\beta\gamma} \bigr) + \bigl( \Lambda_{\gamma\alpha} - d \log g_{\gamma\alpha} \bigr) = 0\\
	&\delta \, \bigl\{ \Lambda_{\alpha\beta} - d \log g_{\alpha\beta} \bigr\} = 0
\end{split}\]
and, since the sheaf of 1-forms is fine, hence acyclic, we obtain:
	\[\Lambda_{\alpha\beta} - d \log g_{\alpha\beta} = A_{\alpha} - A_{\beta} \; .
\]

\paragraph{}We now define the integral of the connection. For a fixed $\Gamma \in \Sigma X$, there exists $(\tau, \varphi) \in J$ such that $\gamma \in V_{(\tau, \varphi)}$. We define:
\begin{equation}\label{WilsonLineGerbe}
	\int_{\Gamma} B \;:=\; \sum_{(a,b,c) \in T_{\tau}} \; \biggl( \int_{\Gamma(\sigma^{2}_{(a,b,c)})} B_{\varphi(a,b,c)} + \int_{\Gamma(\partial \sigma^{2}_{(a,b,c)})} A_{\varphi(a,b,c)} \biggr) \; .
\end{equation}
As before, the logarithm can be taken since we have chosen a good cover and it is  defined up to $2\pi i \,\mathbb{Z}$, so that $\exp\bigl(2\pi i \cdot \int_{\Gamma} B\bigr)$ is as well-defined number. The contribution of $A$ to the internal edges cancel in pairs, so only the integral of $A$ on boundary terms remains. That is why this expression is usually denoted by:
	\[\int_{\Gamma} B + \oint_{\partial \Gamma} A \; .
\]
This expression is equivalent to the one obtained by changing $B$ choosing transition functions on the boundary corresponding to the fixed realization.

\section{Classification by hypercohomology}

We are now ready to describe the classification group for $B$-field and $A$-field configurations in superstring theory with a single D-brane. Our background is specified in particular by a space-time gerbe $\mathcal{G}$ belonging to the following hypercohomology group\footnote{We refer to appendices \ref{AppHyperC} and \ref{AppGerbes} for notations.}:
\begin{equation}\label{GerbeB}
	\mathcal{G} = [\,\{g_{\alpha\beta\gamma}, -\Lambda_{\alpha\beta}, B_{\alpha}\}\,] \in \check{H}^{2}(X, \, \underline{S}^{1} \overset{\tilde{d}}\longrightarrow \Omega^{1}_{\mathbb{R}} \overset{d}\longrightarrow \Omega^{2}_{\mathbb{R}}\,)
\end{equation}
where $\tilde{d} = (2\pi i)^{-1}\, d \circ \log\,$, $g_{\alpha\beta\gamma}$ are functions from triple intersections to $S^{1}$, $\Lambda_{\alpha\beta}$ are 1-forms on double intersections and $B_{\alpha}$ are 2-forms on the opens sets of the cover. In \eqref{GerbeB}, we denote by $\underline{S}^{1}$ the sheaf of smooth $S^{1}$-valued functions on $X$ and by $\Omega^{p}_{\mathbb{R}}$ the sheaf of real $p$-forms. On a single brane $Y \subset X$ we consider the restriction of the space-time gerbe, for which we use the same notation $\mathcal{G}\,\vert_{Y} = [\,\{g_{\alpha\beta\gamma}, -\Lambda_{\alpha\beta}, B_{\alpha}\}\,] \in \check{H}^{2}(Y, \, \underline{S}^{1} \rightarrow \Omega^{1}_{\mathbb{R}} \rightarrow \Omega^{2}_{\mathbb{R}}\,)$. To give a meaning to the holonomy for open surfaces with boundary on $Y$, we must fix a specific representative of the class $\mathcal{G}\,\vert_{Y}$, i.e., a specific hypercocycle; this operation is analogous to fixing a set of local sections on a line bundle up to pull-back by isomorphism (see section \ref{CocyclesCohomology}). To compensate for the possible non-definiteness of $\pfaff\,D_{\phi}$, this hypercocycle must take the form $\{\eta_{\alpha\beta\gamma}, 0, B+F\}$, with $\eta_{\alpha\beta\gamma}$ representing the class $w_{2} \in H^{2}(Y, S^{1})$, denoting by $S^{1}$ the \emph{constant} sheaf. Here $B+F$ is a 2-form globally defined on $Y$, which we now explain in detail. The choice of the specific cocycle $\eta_{\alpha\beta\gamma}$ in the class $w_{2}$ turns out to be immaterial, as we will show later.

\paragraph{}In order to obtain the hypercocycle $\{\eta_{\alpha\beta\gamma}, 0, B+F\}$ from any gauge representative $\{g_{\alpha\beta\gamma},$ $-\Lambda_{\alpha\beta}, B_{\alpha}\}$ of the gerbe $\mathcal{G}\,\vert_{Y}$, the brane must provide a reparametrization of $\mathcal{G}\,\vert_{Y}$, which, by an active point of view, is a hypercoboundary, i.e., a geometrically trivial gerbe. That is, given $\{g_{\alpha\beta\gamma}, -\Lambda_{\alpha\beta}, B_{\alpha}\}$, the brane must provide a coordinate change $\{g_{\alpha\beta\gamma}^{-1} \cdot \eta_{\alpha\beta\gamma}, \Lambda_{\alpha\beta}, dA_{\alpha}\}$, so that:
\begin{equation}
	\{g_{\alpha\beta\gamma}, -\Lambda_{\alpha\beta}, B_{\alpha}\} \cdot \{g_{\alpha\beta\gamma}^{-1} \cdot \eta_{\alpha\beta\gamma}, \Lambda_{\alpha\beta}, dA_{\alpha}\} = \{\eta_{\alpha\beta\gamma}, 0, B+F\}
\end{equation}
for a globally defined $B + F = B_{\alpha} + dA_{\alpha}$. In order for this correction to be geometrically trivial, it must be that:
\begin{equation}\label{CoboundaryA}
	\{g_{\alpha\beta\gamma}^{-1} \cdot \eta_{\alpha\beta\gamma}, \Lambda_{\alpha\beta}, dA_{\alpha}\} = \check{\delta}^{1}\{h_{\alpha\beta}, A_{\alpha}\}
\end{equation}
i.e. $\{g_{\alpha\beta\gamma}^{-1} \cdot \eta_{\alpha\beta\gamma}, \Lambda_{\alpha\beta}, dA_{\alpha}\} = \{\check{\delta}^{1}h_{\alpha\beta}, -\tilde{d} h_{\alpha\beta} + A_{\beta} - A_{\alpha}, dA_{\alpha}\}$. For this to hold one must have:
\begin{itemize}
	\item $\{g_{\alpha\beta\gamma}^{-1} \cdot \eta_{\alpha\beta\gamma}\} = \{\check{\delta}^{1}h_{\alpha\beta}\}$: this is precisely the statement of Freed-Witten anomaly, since, considering the Bockstein homomorphism $\beta$ in degree 2 of the sequence $0 \rightarrow \mathbb{Z} \rightarrow \underline{\mathbb{R}} \rightarrow \underline{S}^{1} \rightarrow 0$, this is equivalent to $\beta(\,[\,g_{\alpha\beta\gamma}\,]\,) = \beta(\,[\,\eta_{\alpha\beta\gamma}\,]\,)$, i.e., $\zeta\,\vert_{Y} = W_{3}(Y)$; only under this condition is $g_{\alpha\beta\gamma}^{-1} \cdot \eta_{\alpha\beta\gamma}$ trivial in the $\underline{S}^{1}$-cohomology;
	\item $A_{\beta} - A_{\alpha} = \tilde{d} h_{\alpha\beta} + \Lambda_{\alpha\beta}$: these must be the transition relations for $A_{\alpha}$ (coherently with \cite{Kapustin}); this is always possible since $\check{\delta}^{1}\{\tilde{d} h_{\alpha\beta}\} = \{\,\tilde{d}(\,\eta_{\alpha\beta\gamma} - g_{\alpha\beta\gamma}\,)\,\} = \{-\tilde{d} g_{\alpha\beta\gamma} \}= -\check{\delta}^{1}\{\Lambda_{\alpha\beta}\}$ and $\Omega^{1}_{\mathbb{R}}$ is acyclic.
\end{itemize}
From the transition relations of $A_{\alpha}$ we obtain $dA_{\beta} - dA_{\alpha} = d\Lambda_{\alpha\beta}$, thus $B+F$ is globally defined. Of course $B_{\alpha}$ and $A_{\alpha}$ themselves depends on the gauge choices, while $B+F$ is gauge-invariant.\footnote{We remark that, for $W_{3}(Y) = 0$, from the exact sequence $0 \rightarrow \mathbb{Z} \rightarrow \underline{\mathbb{R}} \rightarrow \underline{S}^{1} \rightarrow 0$ it follows that $w_{2}(Y)$, having image $0$ under the degree-2 Bockstein homomorphism, by exactness can be lifted to a real form $G$ on $Y$. Therefore, the gerbe $[\,\{\eta_{\alpha\beta\gamma}, 0, B+F\}\,]$ can be also represented by $[\,\{1,0,B+F+G\}\,]$: however, this is not the cocycle we need, since we need transition function realizing the class $w_{2}(Y)$. These two cocycles are equivalent on closed surfaces, since they represent the same gerbe, but not on open ones.}

\paragraph{}Let us now discuss the role of the representative $\eta_{\alpha\beta\gamma}$ of the class $w_{2}(Y) \in \check{H}^{2}(X, S^{1})$. The choice of a different representative corresponds to changing by \emph{constant} local functions the chosen sections of the bundle over loop space, which define the holonomy for open surfaces. This kind of ambiguity is also present for the Pfaffian, since it also defines a section of a flat bundle with the same holonomy. If $w_{2}(Y) \neq 0$, we have no possibility to eliminate this non-definiteness. We can only choose the sections for the Pfaffian and for the gerbe, in such a way that on the tensor product we have a \emph{global} flat section, up to an immaterial overall constant. Instead, if $w_{2} = 0$, both the pfaffian and the gerbe are geometrically trivial, thus we have a preferred choice, given by a global flat section for both. In this case, we fix the canonical representative $\eta_{\alpha\beta\gamma} = 1$. We will see in the following the consequences of this fact for the gauge theory of the D-brane.

\paragraph{}How can we jointly characterize B-field and A-field taking into account the gauge transformations contained in the previous description? This unifying role is played by a certain hypercohomology group, which we would like now to introduce. Since this construction is not very familiar in the literature, we would like for pedagogical reason to start with the analogous group for line bundles.

\subsection{Line bundles}

Let us consider an embedding of manifolds $i: Y \rightarrow X$: we want to describe the group of line bundles on $X$ which are trivial on $Y$, with a fixed trivialization. We recall that $\underline{S}^{1}$ is the sheaf of smooth functions on $X$: it turns out that the sheaf of smooth functions on $Y$ is its pull-back $i^{*}\underline{S}^{1}$. We thus obtain a cochain map $(i^{*})^{p}: \check{C}^{p}(X, \underline{S}^{1}) \longrightarrow \check{C}^{p}(Y, \underline{S}^{1})$, which can be described as follows: we choose a good cover $\mathfrak{U}$ of $X$ restricting to a good cover $\mathfrak{U} \,\vert_{Y}$ of $Y$, such that every $p$-intersection $U_{i_{0} \cdots i_{p}} \vert_{Y}$ comes from a unique $p$-intersection $U_{i_{0} \cdots i_{p}}$ on $X$. Given a $p$-cochain $\oplus_{i_{0} < \cdots < i_{p}} \, f_{i_{0} \cdots i_{p}}$, we restrict $f_{i_{0} \cdots i_{p}}$ to $U_{i_{0} \cdots i_{p}} \vert_{Y}$ whenever the latter is non-empty. In this way we obtain a double complex:
	\[\xymatrix{
	\check{C}^{0}(Y, \underline{S}^{1}) \ar[r]^{\check{\delta}^{0}} & \check{C}^{1}(Y, \underline{S}^{1}) \ar[r]^{\check{\delta}^{1}} & \check{C}^{2}(Y, \underline{S}^{1}) \ar[r]^{\check{\delta}^{2}} & \cdots\\
	\check{C}^{0}(X, \underline{S}^{1})  \ar[u]^{(i^{*})^{0}} \ar[r]^{\check{\delta}^{0}} & \check{C}^{1}(X, \underline{S}^{1}) \ar[u]^{(i^{*})^{1}} \ar[r]^{\check{\delta}^{1}} & \check{C}^{2}(X, \underline{S}^{1}) \ar[u]^{(i^{*})^{2}} \ar[r]^{\check{\delta}^{2}} & \cdots
	}
\]
We denote by $\check{H}^{\bullet}(X, \underline{S}^{1}, Y)$ the hypercohomology of this double complex. We claim that $\check{H}^{1}(X, \underline{S}^{1}, Y)$ is the group we are looking for. In fact, the latter can be defined in the following way: we choose a line bundle $L$ on $X$ with a fixed set of local sections $\{s_{\alpha}\}$, so that the transition functions are $\{g_{\alpha\beta}\}$ for $g_{\alpha\beta} = s_{\alpha} / s_{\beta}$. We consider $\{s_{\alpha}\,\vert_{Y}\}$ and we express the trivialization by means of local functions $\{f_{\alpha}\}$ on $Y$ such that $f_{\alpha} \cdot s_{\alpha}\,\vert_{Y}$ gives a global section of $L\vert_{Y}$. We have that $\check{C}^{1}(X, \underline{S}^{1}, Y) = \check{C}^{1}(X, \underline{S}^{1}) \oplus \check{C}^{0}(Y, \underline{S}^{1})$, so that we can consider the hypercochain $\{g_{\alpha\beta}, f_{\alpha}\}$. We now claim that this is a hypercocycle: to see this, we describe the cohomology group $\check{H}^{1}(X, \underline{S}^{1}, Y)$.
\begin{itemize}
	\item \emph{Cocycles:} since $\check{\delta}^{1}\{g_{\alpha\beta}, f_{\alpha}\} = \{\check{\delta}^{1}g_{\alpha\beta}, ((i^{*})^{1}g_{\alpha\beta})^{-1} \cdot f_{\beta} f_{\alpha}^{-1}\}$, cocycles are characterized by two conditions: $\check{\delta}^{1}g_{\alpha\beta} = 0$, i.e., $g_{\alpha\beta}$ is a line bundle $L$ on $X$, and $(i^{*})^{1}g_{\alpha\beta} = f_{\beta} f_{\alpha}^{-1}$, i.e., $f_{\alpha}$ trivializes $L\vert_{Y}$.
	\item \emph{Coboundaries:} $\check{\delta}^{0}\{g_{\alpha}\} = \{\check{\delta}^{0}g_{\alpha}, (i^{*})^{0}g_{\alpha}\}$ thus coboundaries represents line bundles which are trivial on $X$, with a trivialization on $X$ restricting to to chosen one on $Y$.
\end{itemize}
To explain the structure of the coboundaries, let us remark that if we choose different sections $\{s'_{\alpha} = \varphi_{\alpha} \cdot s_{\alpha}\}$, the same trivialization is expressed by $f'_{\alpha} = \varphi_{\alpha}\vert_{Y}^{-1} \cdot f_{\alpha}$. Thus the coordinate change is given by $\{\varphi_{\alpha}^{-1}\varphi_{\beta}, \varphi_{\alpha}\vert_{Y}\}$, which can be seen, by an active point of view, as a $X \times \mathbb{C}$ with the trivialization $Y \times \{1\}$ on $Y$, i.e., a trivial bundle with a fixed global section on $X$ restricting to the chosen trivialization on $Y$. Hence, $\check{H}^{1}(X, \underline{S}^{1}, Y)$ is the group we are looking for.

\subsubsection{Line bundles with connection}

Let us now define the analogous group for bundles with connection. The relevant complex is the following:
\[\xymatrix{
	\check{C}^{0}(X, \Omega^{1}_{\mathbb{R}}) \oplus \check{C}^{0}(Y, \underline{S}^{1}) \ar[r]^{\check{\delta}^{0} \oplus \check{\delta}^{0}} & \check{C}^{1}(X, \Omega^{1}_{\mathbb{R}}) \oplus \check{C}^{1}(Y, \underline{S}^{1}) \ar[r]^{\check{\delta}^{1} \oplus \check{\delta}^{1}} & \check{C}^{2}(X, \Omega^{1}_{\mathbb{R}}) \oplus \check{C}^{2}(Y, \underline{S}^{1}) \ar[r]^{\phantom{XXXXXXX} \check{\delta}^{2} \oplus \check{\delta}^{2}} & \cdots\\
	\,\phantom{XX} \check{C}^{0}(X, \underline{S}^{1}) \,\phantom{XX} \ar[u]^{\tilde{d} \,\oplus\, (i^{*})^{2}} \ar[r]^{\check{\delta}^{0}} & \,\phantom{XX} \check{C}^{1}(X, \underline{S}^{1}) \,\phantom{XX} \ar[r]^{\check{\delta}^{1}} \ar[u]^{\tilde{d} \,\oplus\, (i^{*})^{1}} & \,\phantom{XX} \check{C}^{2}(X, \underline{S}^{1}) \,\phantom{XX} \ar[u]^{\tilde{d} \,\oplus\, (i^{*})^{2}} \ar[r]^{\phantom{XXXXXX} \check{\delta}^{2}} & \cdots
	}
\]
We denote by $\check{H}^{\bullet}(X, \underline{S}^{1} \rightarrow \Omega^{1}_{\mathbb{R}}, Y)$ the hypercohomology of this double complex. We claim that the group we are looking for is $\check{H}^{1}(X, \underline{S}^{1} \rightarrow \Omega^{1}_{\mathbb{R}}, Y)$. The cochains are given by $\check{C}^{1}(X, \underline{S}^{1} \rightarrow \Omega^{1}_{\mathbb{R}}, Y) = \check{C}^{1}(X, \underline{S}^{1}) \,\oplus\, \check{C}^{0}(Y, \Omega^{1}_{\mathbb{R}}) \,\oplus\, \check{C}^{0}(Y, \underline{S}^{1})$, so that we consider $\{g_{\alpha\beta}, -A_{\alpha}, f_{\alpha}\}$.
\begin{itemize}
	\item \emph{Cocycles:} since $\check{\delta}^{1}\{g_{\alpha\beta}, -A_{\alpha}, f_{\alpha}\} = \{\check{\delta}^{1}g_{\alpha\beta}, -\tilde{d} g_{\alpha\beta} - A_{\beta} + A_{\alpha}, ((i^{*})^{1}g_{\alpha\beta})^{-1} \cdot f_{\beta} f_{\alpha}^{-1}\}$, cocycles are characterized by three conditions: $\check{\delta}^{1}g_{\alpha\beta} = 0$, i.e., $g_{\alpha\beta}$ is a line bundle $L$ on $X$, $A_{\alpha} - A_{\beta} = \tilde{d}g_{\alpha\beta}$, i.e., $A_{\alpha}$ is a connection on $L$, and $(i^{*})^{1}g_{\alpha\beta} = f_{\beta} f_{\alpha}^{-1}$, i.e., $f_{\alpha}$ trivializes $L\vert_{Y}$.
	\item \emph{Coboundaries:} since $\check{\delta}^{0}\{g_{\alpha}\} = \{\check{\delta}^{0}g_{\alpha}, \tilde{d}g_{\alpha\beta}, (i^{*})^{0}g_{\alpha}\}$, coboundaries represents line bundles which are geometrically trivial on $X$ (see appendix), with a trivialization on $X$ restricting to the chosen one on $Y$.
\end{itemize}

\subsection{Gerbes}

Let us now define the analogous group for gerbes with connection. The relevant complex is the following\footnote{The maps denoted by matrices are supposed to multiply from the right the row vector in the domain.}:
\[\xymatrix{
	\check{C}^{0}(X, \Omega^{2}_{\mathbb{R}}) \oplus \check{C}^{0}(Y, \Omega^{1}_{\mathbb{R}}) \ar[r]^{\check{\delta}^{0} \oplus \check{\delta}^{0}} & \check{C}^{1}(X, \Omega^{2}_{\mathbb{R}}) \oplus \check{C}^{1}(Y, \Omega^{1}_{\mathbb{R}}) \ar[r]^{\check{\delta}^{1} \oplus \check{\delta}^{1}} & \check{C}^{2}(X, \Omega^{2}_{\mathbb{R}}) \oplus \check{C}^{2}(Y, \Omega^{1}_{\mathbb{R}}) \ar[r]^{\phantom{XXXXXXX} \check{\delta}^{2} \oplus \check{\delta}^{2}} & \cdots\\ \\
	\check{C}^{0}(X, \Omega^{1}_{\mathbb{R}}) \oplus \check{C}^{0}(Y, \underline{S}^{1}) \ar[uu]_{\footnotesize{\begin{bmatrix} d & (i^{*})^{0} \\ 0 & -\tilde{d} \end{bmatrix}}} \ar[r]^{\check{\delta}^{0} \oplus \check{\delta}^{0}} & \check{C}^{1}(X, \Omega^{1}_{\mathbb{R}}) \oplus \check{C}^{1}(Y, \underline{S}^{1}) \ar[r]^{\check{\delta}^{1} \oplus \check{\delta}^{1}} \ar[uu]_{\footnotesize{\begin{bmatrix} d & (i^{*})^{1} \\ 0 & -\tilde{d} \end{bmatrix}}} & \check{C}^{2}(X, \Omega^{1}_{\mathbb{R}}) \oplus \check{C}^{2}(Y, \underline{S}^{1}) \ar[uu]_{\footnotesize{\begin{bmatrix} d & (i^{*})^{2} \\ 0 & -\tilde{d} \end{bmatrix}}} \ar[r]^{\phantom{XXXXXXX} \check{\delta}^{2} \oplus \check{\delta}^{2}} & \cdots\\ \\
	\check{C}^{0}(X, \underline{S}^{1}) \ar[uu]_{\tilde{d} \,\oplus\, (i^{*})^{0}} \ar[r]^{\check{\delta}^{0}} & \check{C}^{1}(X, \underline{S}^{1}) \ar[uu]_{\tilde{d} \,\oplus\, (i^{*})^{1}} \ar[r]^{\check{\delta}^{1}} & \check{C}^{2}(X, \underline{S}^{1}) \ar[uu]_{\tilde{d} \,\oplus\, (i^{*})^{2}} \ar[r]^{\phantom{XXXXX} \check{\delta}^{2}} & \cdots
	}
\]
We denote by $\check{H}^{\bullet}(X, \underline{S}^{1} \rightarrow \Omega^{1}_{\mathbb{R}} \rightarrow \Omega^{2}_{\mathbb{R}}, Y)$ the hypercohomology of this double complex. We claim that the group we are looking for is $\check{H}^{2}(X, \underline{S}^{1} \rightarrow \Omega^{1}_{\mathbb{R}} \rightarrow \Omega^{2}_{\mathbb{R}}, Y)$. The cochains are given by $\check{C}^{2}(X, \underline{S}^{1} \rightarrow \Omega^{1}_{\mathbb{R}} \rightarrow \Omega^{2}_{\mathbb{R}}, Y) = \check{C}^{2}(X, \underline{S}^{1}) \,\oplus\, \check{C}^{1}(X, \Omega^{1}_{\mathbb{R}}) \,\oplus\, \check{C}^{1}(Y, \underline{S}^{1}) \,\oplus\, \check{C}^{0}(X, \Omega^{2}_{\mathbb{R}}) \,\oplus\, \check{C}^{0}(Y, \Omega^{1}_{\mathbb{R}})$, so that we consider $\{g_{\alpha\beta\gamma}, -\Lambda_{\alpha\beta}, h_{\alpha\beta}, B_{\alpha}, -A_{\alpha}\}$.
\begin{itemize}
	\item \emph{Cocycles:} since $\check{\delta}^{2}\{g_{\alpha\beta\gamma}, -\Lambda_{\alpha\beta}, h_{\alpha\beta}, B_{\alpha}, -A_{\alpha}\} = \{\check{\delta}^{2}g_{\alpha\beta\gamma}, \tilde{d}g_{\alpha\beta\gamma} + \check{\delta}^{1}(-\Lambda_{\alpha\beta}), (i^{*})^{2}g_{\alpha\beta\gamma} \cdot \check{\delta}^{2}h_{\alpha\beta},$ $-d(-\Lambda_{\alpha\beta}) + B_{\beta} - B_{\alpha}, -(i^{*})^{1}(-\Lambda_{\alpha\beta}) + \tilde{d}h_{\alpha\beta} + A_{\alpha} - A_{\beta}\}$, cocycles are characterized exactly by the condition we need in order for $\{g_{\alpha\beta\gamma}, -\Lambda_{\alpha\beta}, B_{\alpha}\}$ to be a gerbe with connection and $\{h_{\alpha\beta}, A_{\alpha}\}$ to trivialize it on $Y$;
	\item \emph{Coboundaries:} since $\check{\delta}^{1}\{g_{\alpha\beta}, \Lambda_{\alpha}, h_{\alpha}\} = \{\check{\delta}^{1}g_{\alpha\beta}, -\tilde{d}g_{\alpha\beta} + \Lambda_{\beta} - \Lambda_{\alpha}, ((i^{*})^{1}g_{\alpha\beta})^{-1} \cdot h_{\beta} h_{\alpha}^{-1}, d\Lambda_{\alpha},$ $(i^{*})^{0} \Lambda_{\alpha} - \tilde{d}h_{\alpha}\}$, coboundaries represent gerbes which are geometrically trivial on $X$ (see appendix), with a trivialization on $X$ restricting to the chosen one on $Y$.
\end{itemize}

\paragraph{}There is a last step to obtain the classifying set of $B$-field and $A$-field configurations: in general we do not ask for a trivialization of the gerbe on $Y$, but for a cocycle whose transition functions represent the class $w_{2}(Y) \in H^{2}(Y, S^{1})$. The transition functions of a coboundary in the previous picture represent the zero class, so they are consistent only for $w_{2}(Y) = 0$. Hence, we cannot consider the hypercohomology group, but one of its cosets in the group of \emph{cochains} up to coboundaries. In fact, the condition we need is not cocycle condition, but:
\begin{equation}\label{Coset}
	\check{\delta}^{2}\{g_{\alpha\beta\gamma}, -\Lambda_{\alpha\beta}, h_{\alpha\beta}, B_{\alpha}, -A_{\alpha}\} = \{0, 0, \eta_{\alpha\beta\gamma}, 0, 0\}
\end{equation}
thus we need the coset made by cochains satisfying \eqref{Coset} up to coboundaries. Actually, we need anyone of these cosets for $[\,\{\,\eta_{\alpha\beta\gamma}\,\}\,] = w_{2}(Y) \in \check{H}^{2}(Y, S^{1})$. We denote their union by:
\begin{equation}
	\check{H}^{2}_{w_{2}(Y)}(X, \underline{S}^{1} \rightarrow \Omega^{1}_{\mathbb{R}} \rightarrow \Omega^{2}_{\mathbb{R}}, Y)
\end{equation}
and this is the set of configurations we are looking for.

\section{Gauge theory on a single D-brane}\label{GaugeTheories}

We are now ready to discuss the possible geometric structures of the gauge theory on the D-brane, arising from the previous picture. The main distinction turns out to be whether or not the B-field is flat when restricted to the D-brane.

\subsection{Generic $B$-field}

We consider the coordinate change given by the D-brane:
\begin{equation}\label{CoordChange}
\begin{split}
	\{g_{\alpha\beta\gamma}, &-\Lambda_{\alpha\beta}, B_{\alpha}\} \cdot \{g_{\alpha\beta\gamma}^{-1} \cdot \eta_{\alpha\beta\gamma}, \Lambda_{\alpha\beta}, dA_{\alpha}\} = \{\eta_{\alpha\beta\gamma}, 0, B+F\}\\
	&\{g_{\alpha\beta\gamma}^{-1} \cdot \eta_{\alpha\beta\gamma}, \Lambda_{\alpha\beta}, dA_{\alpha}\} = \{\check{\delta}^{1}h_{\alpha\beta}, -\tilde{d}h_{\alpha\beta} + A_{\beta} - A_{\alpha}, dA_{\alpha}\} \; .
\end{split}
\end{equation}
Since, by Freed-Witten anomaly, $[\,\{\,g_{\alpha\beta\gamma}\,\}\,] = [\,\{\,\eta_{\alpha\beta\gamma}\,\}\,] \in \check{H}^{2}(Y, \underline{S}^{1})$ (not the constant sheaf $S^{1}$, the sheaf of functions $\underline{S}^{1}$), we can always choose a gauge $\{\eta_{\alpha\beta\gamma}, 0, B\}$, but we can also consider any gauge $\{\eta_{\alpha\beta\gamma}, 0, B'\}$ with $B'-B$ a closed form representing an integral de Rham class: for a bundle, this corresponds to the free choice of a global automorphism.\footnote{For gerbes, we directly see this from the fact that $(1, 0, \Phi)$ is a hypercoboundary for $\Phi$ integral. Indeed, we have:
	\[\Phi\vert_{U_{\alpha}} = d\varphi_{\alpha} \qquad \varphi_{\beta} - \varphi_{\alpha} = d\rho_{\alpha\beta} \qquad \rho_{\alpha\beta} + \rho_{\beta\gamma} + \rho_{\gamma\alpha} = c_{\alpha\beta\gamma} \in \mathbb{Z}
\]
thus $\varphi_{\beta} - \varphi_{\alpha} = \tilde{d}h_{\alpha\beta}$ for $h_{\alpha\beta} = \exp(2\pi i \cdot \rho_{\alpha\beta})$ and $\check{\delta}^{1}h_{\alpha\beta} = 1$. Hence, $(1, 0, \Phi) = \check{\delta}^{1}(h_{\alpha\beta}, \varphi_{\alpha})$.} Given a certain gauge of the form $\{\eta_{\alpha\beta\gamma}, 0, B\}$, the brane gives a correction $\{1, 0, F\}$ to arrive at the fixed gauge $\{\eta_{\alpha\beta\gamma}, 0, B+F\}$. In fact, \eqref{CoordChange} becomes:
\begin{equation}
\begin{split}
	\{\eta_{\alpha\beta\gamma}, &0, B\} \cdot \{1, 0, dA_{\alpha}\} = \{\eta_{\alpha\beta\gamma}, 0, B+F\}\\
	&\{1, 0, dA_{\alpha}\} = \{\check{\delta}^{1}h_{\alpha\beta}, -\tilde{d}h_{\alpha\beta} + A_{\beta} - A_{\alpha}, dA_{\alpha}\} \; .
\end{split}
\end{equation}
We thus get $\check{\delta}^{1}h_{\alpha\beta} = 1$ and $-\tilde{d}h_{\alpha\beta} + A_{\beta} - A_{\alpha} = 0$, so $h_{\alpha\beta}$ give a gauge bundle on the brane with connection $-A_{\alpha}$ and Chern class $[\,-F\,]$. However, since $B$ and $F$ are arbitrary, such bundle is defined up to large gauge transformations $B \rightarrow B + \Phi$ and $F \rightarrow F - \Phi$ for $\Phi$ integral.\footnote{In particular, we can always choose the gauge $F = 0$, obtaining a flat line bundle.}

\paragraph{}Moreover, we have the freedom to choose a different representative $\eta_{\alpha\beta\gamma} \cdot \check{\delta}^{1}\lambda_{\alpha\beta}$ of $w_{2}(Y) \in \check{H}^{2}(Y, S^{1})$. This is equivalent to consider:
\begin{equation}
\begin{split}
	\{\eta_{\alpha\beta\gamma}, &0, B\} \cdot \{\check{\delta}\lambda_{\alpha\beta}, 0, dA_{\alpha}\} = \{\eta_{\alpha\beta\gamma} \cdot \check{\delta}\lambda_{\alpha\beta}, 0, B+F\}\\
	&\{\check{\delta}\lambda_{\alpha\beta}, 0, dA_{\alpha}\} = \{\check{\delta}h_{\alpha\beta}, -\tilde{d}h_{\alpha\beta} + A_{\beta} - A_{\alpha}, dA_{\alpha}\} \; .
\end{split}
\end{equation}
We thus obtain that $\check{\delta}h_{\alpha\beta} = \check{\delta}\lambda_{\alpha\beta}$, i.e., $\check{\delta}(h_{\alpha\beta} / \lambda_{\alpha\beta}) = 1$. So, instead of $\{h_{\alpha\beta}\}$, we consider the bundle $[\,h_{\alpha\beta} / \lambda_{\alpha\beta}\,]$ instead of $[\,h_{\alpha\beta}\,]$. Since the functions $\lambda_{\alpha\beta}$ are constant, the real image of the Chern class is the same. In fact, if we write $h_{\alpha\beta} = \exp(2\pi i \cdot \tilde{h}_{\alpha\beta})$ and $\lambda_{\alpha\beta} = \exp(2\pi i \cdot \tilde{\lambda}_{\alpha\beta})$, we have that $\tilde{h}_{\alpha\beta} + \tilde{h}_{\beta\gamma} + \tilde{h}_{\gamma\alpha} = \tilde{h}_{\alpha\beta\gamma} \in \mathbb{Z}$ defining the first Chern class, and similarly $\tilde{\lambda}_{\alpha\beta} + \tilde{\lambda}_{\beta\gamma} + \tilde{\lambda}_{\gamma\alpha} = \tilde{\lambda}_{\alpha\beta\gamma} \in \mathbb{Z}$. However, since $\tilde{\lambda}_{\alpha\beta}$ are constant, $\tilde{\lambda}_{\alpha\beta\gamma}$ is a coboundary in the sheaf $\mathbb{R}$ and the real image of the Chern class of $\lambda_{\alpha\beta}$ is $0$.

This means that we fix a line bundle \emph{up to the torsion part}. Thus, the holonomy of $-A_{\alpha}$ is defined also up to the torsion part: this ambiguity is compensated for by the one of the pfaffian, due to the need of obtaining a global section of the tensor product. If $w_{2} = 0$, we can choose the preferred representative $\eta_{\alpha\beta\gamma} = 1$, thus we completely fix a line bundle up to large gauge transformation.

\subsection{Flat $B$-field}

If $B$ is flat, its holonomy is a class $\Hol(B\vert_{Y}) \in H^{2}(Y, S^{1})$ (constant sheaf $S^{1}$). We distinguish three cases:
\begin{itemize}
	\item $\Hol(B\vert_{Y}) = w_{2}(Y) = 0$: as before, we can choose the gauge $\eta_{\alpha\beta\gamma} = 1$, but, via an operation analogous to choosing \emph{parallel} local sections for line bundles, we can obtain $\{1, 0, 0\}$ instead of a generic $\{1, 0, B\}$. The choice $B = 0$ is canonical (it fixes also large gauge transformations). Thus we get $\{1, 0, 0\} \,\cdot\, \{1, 0, dA_{\alpha}\} = \{1, 0, F\}$ with $\{1, 0, dA_{\alpha}\} = \{\check{\delta}^{1}h_{\alpha\beta}, -\tilde{d}h_{\alpha\beta} + A_{\beta} - A_{\alpha}, dA_{\alpha}\}$. Hence we have $\check{\delta}^{1}h_{\alpha\beta} = 1$ and $A_{\beta} - A_{\alpha} = \tilde{d}_{\alpha\beta}$. In this case, we obtain a line bundle $L$ with connection $-A_{\alpha}$ and Chern class $c_{1}(L)$ such that $c_{1}(L) \otimes_{\mathbb{Z}} \mathbb{R} = [\,-F\,]_{dR}$, i.e., a gauge theory in the usual sense, canonically fixed. However, we will see in the following that, also in this case, there is a residual freedom in the choice of the bundle.
	\item $\Hol(B\vert_{Y}) = w_{2}(Y)$: as before, we choose $\{\eta_{\alpha\beta\gamma}, 0, 0\}$ instead of a generic $\{\eta_{\alpha\beta\gamma}, 0, B\}$. The choice $B = 0$ is canonical (it fixes also large gauge transformations). Thus we get $\{\eta_{\alpha\beta\gamma}, 0, 0\} \cdot \{1, 0, dA_{\alpha}\} = \{\eta_{\alpha\beta\gamma}, 0, F\}$ with $\{1, 0, dA_{\alpha}\} = \{\check{\delta}^{1}h_{\alpha\beta}, -\tilde{d}h_{\alpha\beta} + A_{\beta} - A_{\alpha}, dA_{\alpha}\}$, or, as discussed before, $\{\eta_{\alpha\beta\gamma}, 0, 0\} \cdot \{\check{\delta}^{1}\lambda_{\alpha\beta}, 0, dA_{\alpha}\} = \{\eta_{\alpha\beta\gamma} \cdot \check{\delta}^{1}\lambda_{\alpha\beta}, 0, F\}$ with $\{\check{\delta}^{1}\lambda_{\alpha\beta}, 0, dA_{\alpha}\} = \{\check{\delta}^{1}h_{\alpha\beta}, -\tilde{d}h_{\alpha\beta} + A_{\beta} - A_{\alpha}, dA_{\alpha}\}$. In this case, we obtain a canonical line bundle with connection $-A_{\alpha}$ up to the torsion part, with real image of the Chern class $[\,-F\,]$.
	\item \emph{$\Hol(B\vert_{Y})$ generic:} in this case, we can use the same picture as for non-flat $B$-fields, obtaining a non-canonical gauge bundle, or we can use flatness to obtain a canonical gauge theory of different nature. In the latter case, we fix a cocycle $\{g_{\alpha\beta\gamma}\}$ such that $[\,\{g_{\alpha\beta\gamma}\}\,] = \Hol(B\vert_{Y}) \in H^{2}(Y, S^{1})$. We thus get a preferred gauge $\{g_{\alpha\beta\gamma}, 0, 0\}$, so that \eqref{CoboundaryA} becomes $\{g_{\alpha\beta\gamma}^{-1} \cdot \eta_{\alpha\beta\gamma}, 0, dA_{\alpha}\} = \{\check{\delta}^{1}h_{\alpha\beta}, -\tilde{d}h_{\alpha\beta} + A_{\beta} - A_{\alpha}, dA_{\alpha}\}$. We obtain $\check{\delta}^{1}h_{\alpha\beta} = g_{\alpha\beta\gamma}^{-1} \cdot \eta_{\alpha\beta\gamma}$ and $A_{\beta} - A_{\alpha} = \tilde{d}h_{\alpha\beta}$. Since $g_{\alpha\beta\gamma}^{-1} \cdot \eta_{\alpha\beta\gamma}$ are constant, we obtain a ``bundle with not integral Chern class'', as explained in the next section.
\end{itemize}

\paragraph{Remark:} We have said above that only for $\Hol(B\vert_{Y}) = 0$ and $w_{2}(Y) = 0$ we are able to recover the torsion of the gauge bundle. Actually, we can still recover the torsion part even if $w_{2}(Y) = 0$ and $B$ is flat. In fact, also in this case we can choose $\eta_{\alpha\beta\gamma} = 1$ fixing the transition function $h_{\alpha\beta}$ of the bundle. Let us consider a fractional bundle $L$ such that $\check{\delta}\{h_{\alpha\beta}\} = \{g_{\alpha\beta\gamma}^{-1}\}$ for $[\,g_{\alpha\beta\gamma}\,] = \Hol(B\vert_{Y}) \in H^{2}(Y, S^{1})$. Then, evaluating the holonomy of $B$ over the generators of $H_{2}(Y, \mathbb{Z})$, we can find a discrete subgroup $\Gamma \leq \mathbb{R}$ such that $c_{1}(L) \in H^{2}(Y, \Gamma)$, so that $c_{1}(L)$ has a torsion part. This is more interesting if we know the fractionality of the brane (see below): for example, if we have a $\frac{1}{n}\,$-fractional gauge theory (e.g., fractional branes from $\mathbb{Z}_{n}$-orbifolds), we have $c_{1}(L) \in H^{2}(Y, \frac{1}{n}\mathbb{Z}) \simeq H^{2}(Y, \mathbb{Z})$.\\$\square$

\paragraph{}A comment is in order when $\Hol(B\vert_{Y}) = w_{2}(Y) = 0$: also in this case, the bundle is not completely fixed, but there is a residual gauge freedom. In fact, such configuration is described by $[\,\{g_{\alpha\beta\gamma}, -\Lambda_{\alpha\beta}, h_{\alpha\beta}, B_{\alpha}, -A_{\alpha}\}\,] \in \check{H}^{2}(X, \underline{S}^{1} \rightarrow \Omega^{1}_{\mathbb{R}} \rightarrow \Omega^{2}_{\mathbb{R}}, Y)$ such that $[\,\{g_{\alpha\beta\gamma}, -\Lambda_{\alpha\beta}, B_{\alpha}\}\,]$ is geometrically trivial on $Y$. As we said, we can choose on $Y$ the preferred gauge $\{1, 0, h_{\alpha\beta}, 0, -A_{\alpha}\}$ so that the cocycle condition gives exactly $\{1, 0, \check{\delta}^{2}h_{\alpha\beta}, 0, \tilde{d}h_{\alpha\beta} + A_{\alpha} - A_{\beta}\} = 0$, i.e. $-A_{\alpha}$ is a connection on the bundle $[\,h_{\alpha\beta}\,]$. There is still a question: how are the possible representatives $\{1, 0, h_{\alpha\beta}, 0, -A_{\alpha}\}$ of the same class? Can they all be obtained via a reparametrization of the bundle $[\,h_{\alpha\beta}, A_{\alpha}\,] \in \check{H}^{1}(Y, \underline{S}^{1} \rightarrow \Omega^{1}_{\mathbb{R}})$? The possible reparametrization are given by:
	\[\begin{split}
	\{1, 0, h_{\alpha\beta}, 0, -A_{\alpha}\} \,\cdot\, \{\check{\delta}^{1}g_{\alpha\beta}, -\tilde{d}g_{\alpha\beta} + \Lambda_{\beta} - \Lambda_{\alpha}, &((i^{*})^{1}g_{\alpha\beta})^{-1} \cdot h_{\beta} h_{\alpha}^{-1}, d\Lambda_{\alpha}, (i^{*})^{0} \Lambda_{\alpha} - \tilde{d}h_{\alpha}\}\\
	&= \{1, 0, h'_{\alpha\beta}, 0, -A'_{\alpha}\}
\end{split}\]
thus we get the conditions:
\begin{equation}\label{ResidualGauge}
	\check{\delta}^{1}g_{\alpha\beta} = 1 \qquad -\tilde{d}g_{\alpha\beta} + \Lambda_{\beta} - \Lambda_{\alpha} = 0 \qquad d\Lambda_{\alpha} = 0 \;.
\end{equation}
If we choose $g_{\alpha\beta} = 1$ and $\Lambda_{\alpha} = 0$ we simply get $h'_{\alpha\beta} = h_{\alpha\beta} \cdot h_{\beta} h_{\alpha}^{-1}$ and $A'_{\alpha} = A_{\alpha} + \tilde{d}h_{\alpha}$, i.e., a reparametrization of $[\,h_{\alpha\beta}, A_{\alpha}\,] \in \check{H}^{1}(Y, \underline{S}^{1} \rightarrow \Omega^{1}_{\mathbb{R}})$, and that is what we expected. But what happens in general? Equations \eqref{ResidualGauge} represent any line bundle $g_{\alpha\beta}$ on the whole space-time $X$ with \emph{flat} connection $-\Lambda_{\alpha}$, thus they represent a residual gauge freedom in the choice of the line bundle over $Y$: \emph{any flat bundle on $Y$ which is the restriction of a \emph{flat} line bundle over $X$ is immaterial for the gauge theory on the D-brane}. Can we give a physical interpretation of this fact?

Let us consider a line bundle $L$ over $Y$ with connection $-A_{\alpha}$: it determines the holonomy as a function from the loop space of $Y$ to $S^{1}$. Actually, we are not interested in a generic loop: we always work with $\partial \Sigma$, with $\Sigma$ in general not contained in $Y$: thus, such loops are in general not homologically trivial on $Y$, but they are so on $X$. Let us suppose that $L$ extends to $\tilde{L}$ over $X$: in this case, we can equally consider the holonomy over $\partial \Sigma$ with respect to $\tilde{L}$. If $\tilde{L}$ is flat, such holonomy becomes an $S^{1}$-cohomology class evaluated over a contractible loop, thus it is $0$. Hence, a bundle extending to a flat one over $X$ gives no contribution to the holonomy over the possible boundaries of the world-sheets. Therefore, also in the case $\Hol(B\vert_{Y}) = w_{2}(Y) = 0$, \emph{we do not have a canonically fixed bundle with connection} on the brane: we rather have an equivalence class of bundles defined up to flat ones extending to flat space-time bundles. For another important comment on this point, see the conclusions.

\section{Real Chern classes}

In the previous section we showed that for $B$ flat we obtain a gauge theory on a generalized bundle: while bundles are represented by cocycles $\{g_{\alpha\beta}\}$ in $\rm\check{C}$ech cohomology, such generalized bundles are represented by cochains whose coboundary $\check{\delta}^{1}\{g_{\alpha\beta}\}$ is made by constant functions (not necessarily $1$), realizing a class in $\check{H}^{2}(X, S^{1})$. We now see that even in these cases we can define connections and first Chern class, but the latter turns out to be any closed form, not necessarily integral.

\paragraph{}Let us consider the definition of Chern class of a trivial bundle: we have a bundle $[\,\{g_{\alpha\beta}\}\,] \in \check{H}^{1}(\mathfrak{U}, \underline{S}^{1})$, so that $g_{\alpha\beta} \cdot g_{\beta\gamma} \cdot g_{\gamma\alpha} = 1$; if $g_{\alpha\beta} = e^{2\pi i \cdot \rho_{\alpha\beta}}$, we have $\rho_{\alpha\beta} + \rho_{\beta\gamma} + \rho_{\gamma\alpha} = \rho_{\alpha\beta\gamma} \in \mathbb{Z}$, so that we obtain a class $[\,\{\rho_{\alpha\beta\gamma}\}\,] \in \check{H}^{2}(\mathfrak{U}, \mathbb{Z})$ which is the first Chern class.

Let us call $\Gamma_{n}$ the subgroup of $S^{1}$ given by the $n$-th root of unity. If we call $\frac{1}{n}\mathbb{Z}$ the subgroup of $\mathbb{R}$ made by the fractions $\frac{k}{n}$ for $k \in \mathbb{Z}$, then $\Gamma_{n} = e^{2\pi i \cdot \frac{1}{n}\mathbb{Z}}$. Let us suppose we have a cochain $\{g_{\alpha\beta}\} \in \check{C}^{1}(\mathfrak{U}, \underline{S}^{1})$ such that $g_{\alpha\beta} \cdot g_{\beta\gamma} \cdot g_{\gamma\alpha} = g_{\alpha\beta\gamma} \in \Gamma_{n}$. Then, for $g_{\alpha\beta} = e^{2\pi i \cdot \rho_{\alpha\beta}}$, we have that $\rho_{\alpha\beta} + \rho_{\beta\gamma} + \rho_{\gamma\alpha} = \rho_{\alpha\beta\gamma} \in \frac{1}{n}\mathbb{Z}$, so that we obtain a rational class $c_{1} = [\,\{\rho_{\alpha\beta\gamma}\}\,] \in \check{H}^{2}(\mathfrak{U}, \mathbb{Q})$ such that $n \cdot c_{1}$ is an integral class. Can we give a geometric interpretation of these classes?

\paragraph{}A 2-cochain can be thought of as a trivialization of a trivialized gerbe, in the same way as a 1-cochain (i.e., a set of local functions) is a trivialization of a trivialized line bundle; thus a line bundle is a trivialization of a gerbe represented by the coboundary $1$, in the same way as a global function is a global section of $X \times \mathbb{C}$. We describe first the easier case of local functions trivializing a line bundle, i.e., we lower by 1 the degree in cohomology. 

\subsection{Trivializations of line bundles}

\subsubsection{Definition}

As line bundles, which are classes in $\check{H}^{1}(\mathfrak{U}, \underline{S}^{1})$, are trivializations of gerbes represented by the coboundary $1$, likewise a section of a line bundle, represented by transition functions equal to 1, is a class in $\check{H}^{0}(\mathfrak{U}, \underline{S}^{1})$, i.e., a function $f: X \rightarrow S^{1}$. A cochain $\{f_{\alpha}\} \in \check{C}^{0}(\mathfrak{U}, \underline{S}^{1})$ is a section of a trivial bundle represented by transition functions $f_{\alpha}^{-1} \cdot f_{\beta}$.

Given a function $f: X \rightarrow S^{1}$, we can naturally define a Chern class $c_{1}(f) \in H^{1}(\mathfrak{U}, \mathbb{Z})$, which is the image under the Bockstein map of $f = [\,\{f_{\alpha}\}\,] \in \check{H}^{0}(\mathfrak{U}, \underline{S}^{1})$. We directly compute it as for bundles: since $f_{\beta} \cdot f_{\alpha}^{-1} = 1$, for $f_{\alpha} = e^{2\pi i \cdot \rho_{\alpha}}$ we have $\rho_{\beta} - \rho_{\alpha} = \rho_{\alpha\beta} \in \mathbb{Z}$, so that we can define a class $c_{1}(f) = [\,\{\rho_{\alpha\beta}\}\,] \in \check{H}^{1}(\mathfrak{U}, \mathbb{Z})$. The geometric interpretation is very simple: $c_{1}(f)$ is the pull-back under $f$ of the generator of $H^{1}(S^{1}, \mathbb{Z}) \simeq \mathbb{Z}$. As we have done for bundles, let us suppose we have a cochain $[\,\{f_{\alpha}\}\,] \in \check{C}^{0}(\mathfrak{U}, \underline{S}^{1})$ such that $f_{\alpha}^{-1} \cdot f_{\beta} = f_{\alpha\beta} \in \Gamma_{n}$. Then $\rho_{\beta} - \rho_{\alpha} = \rho_{\alpha\beta} \in \frac{1}{n} \mathbb{Z}$. Therefore we obtain a class $c_{1} = [\,\{\rho_{\alpha\beta}\}\,] \in \check{H}^{1}(\mathfrak{U}, \mathbb{Q})$ such that $n \cdot c_{1}$ is an integral class.

\paragraph{}From the exact sequences point of view, the Chern class is the image of the Bockstein map of the sequence:
	\[0 \longrightarrow \mathbb{Z} \longrightarrow \underline{\mathbb{R}} \overset{e^{2\pi i \, \cdot}}\longrightarrow \underline{S}^{1} \longrightarrow 0 \; .
\]
In the fractional case, since $\check{\delta}^{0}f_{\alpha}$ takes values in $\Gamma_{n}$, the cochain $\{f_{\alpha}\}$ is a cocycle in $\underline{S}^{1} / \,\Gamma_{n}$. Thus, we consider the sequece:
	\[0 \longrightarrow \textstyle \frac{1}{n} \displaystyle \mathbb{Z} \longrightarrow \underline{\mathbb{R}} \overset{\pi_{\Gamma_{n}} \,\circ\, e^{2\pi i \, \cdot}}\longrightarrow \underline{S}^{1} / \,\Gamma_{n} \longrightarrow 0
\]
and the image of the Bockstein map is exactly the fractional Chern class. We have constructed in this way rational Chern classes, but this is generalizable to any real Chern class. In fact, it is sufficient that $\rho_{\alpha\beta}$ be constant for every $\alpha, \beta$ to apply the previous construction, using the constant sheaf $S^{1}$ instead of $\Gamma_{n}$. The corresponding sequence, which contains all the previous ones by inclusion, is:
	\[0 \longrightarrow \textstyle \mathbb{R} \longrightarrow \underline{\mathbb{R}} \overset{\pi_{S^{1}} \,\circ\, e^{2\pi i \, \cdot}}\longrightarrow \underline{S}^{1} / \,S^{1} \longrightarrow 0 \; .
\]
In other words, if the cochain is a cocycle up to constant functions, we obtain a real Chern class. If these constant functions belong to $\Gamma_{n}$, we obtain a rational Chern class in $\frac{1}{n}\mathbb{Z}$. We now want to give a geometric interpretation of these classes.

\subsubsection{Geometric interpretation}

If we think of the cochain as a trivialization of $X \times \mathbb{C}$, it follows that different trivializations have different Chern classes, depending on the realization of the trivial bundle as $\rm\check{C}$ech coboundary. This seems quite unnatural from a topological point of view, since the particular trivialization should not play any role. However, if we fix a flat connection, we can distinguish a particular class of trivializations, which are parallel with respect to such a connection.

Let us consider a trivial line bundle with a global section and a flat connection $\nabla$, which we think of as $X \times \mathbb{C}$ with a globally defined form $A$, expressing $\nabla$ with respect to the global section $X \times \{1\}$. We know the following facts:
\begin{itemize}
	\item if we choose parallel sections $\{f_{\alpha}\}$, we obtain a trivialization with a real Chern class $c_{1}(\{f_{\alpha}\}) \in \check{H}^{1}(X, \mathbb{R})$, and the local expression of the connection becomes $\{0\}$;
	\item the globally defined connection $A$, expressed with respect to $1$, is closed by flatness, thus it determines a de Rham cohomology class $[\,A\,] \in H^{1}_{dR}(X)$.
\end{itemize}
We now prove that these two classes coincide under the standard isomorphism between $\rm\check{C}$ech and de Rham cohomology. This is the geometric interpretation of real Chern classes: \emph{the real Chern class of a trivialization of $X \times \mathbb{C}$ is the cohomology class of a globally-defined flat connection, expressed with respect to $X \times \{1\}$, for which the trivialization is parallel}.

If the trivial bundle has holonomy $1$ (i.e., \emph{geometrically trivial}), we can find a global parallel section: thus there exists a function $f \in \check{H}^{1}(X, \underline{S}^{1})$ trivializing the bundle, and the Chern class of a function is integral. If we express the connection with respect to $1$ we obtain an integral class $[\,A\,] = [\,f^{-1} df\,]$, while if we express it with respect to the global section $f \cdot 1$ we obtain $0$.

\paragraph{}We now prove the statement. Given $\{f_{\alpha}\} \in \check{C}^{0}(\mathfrak{U}, \underline{S}^{1})$ such that $\check{\delta}^{0}\{f_{\alpha}\} \in \check{C}^{1}(\mathfrak{U}, S^{1})$, we consider the connection $\nabla$ on $X \times \mathbb{C}$ which is represented by $0$ with respect to $\{f_{\alpha}^{-1}\}$. If we represent $\nabla$ with respect to $X \times \{1\}$ we obtain $A_{\alpha} = \tilde{d} f_{\alpha}$, and $A_{\alpha} - A_{\beta} = \tilde{d} (f_{\beta} \cdot f_{\alpha}^{-1} ) = 0$. We thus realize the 1-form $A$ as a $\rm\check{C}$ech cocycle: we have that $A_{\alpha} = (2\pi i)^{-1} d\log f_{\alpha}$ and $(2\pi i)^{-1}\log f_{\beta} - (2\pi i)^{-1}\log f_{\alpha} = (2\pi i)^{-1}\log g_{\alpha\beta} = \rho_{\alpha\beta}$ which is constant, so that $[\,A\,]_{H^{1}_{dR}(X)} \simeq [\,\{\rho_{\alpha\beta}\}\,]_{\check{H}^{1}(X, \mathbb{R})}$. By definition $c_{1}(\{f_{\alpha}\}) = [\,\{\rho_{\alpha\beta}\}\,]$, thus $[\,A\,]_{H^{1}_{dR}(X)} \simeq c_{1}(\{f_{\alpha}\})_{\check{H}^{1}(X, \mathbb{R})}$.

\paragraph{}Moreover, if we consider the sequence $0 \rightarrow \mathbb{Z} \rightarrow \mathbb{R} \rightarrow S^{1} \rightarrow 0$, for $p_{S^{1}}: H^{1}(X, \mathbb{R}) \rightarrow H^{1}(X, S^{1})$, we have that $p_{S^{1}} \, c_{1}(\{f_{\alpha}\}) = p_{S^{1}} \, [\,\rho_{\alpha\beta}\,] = [\,f_{\beta}f_{\alpha}^{-1}\,]_{S^{1}}$. Thus, for $\check{\delta}^{0}\{f_{\alpha}\} \in \check{C}^{1}(X, S^{1})$ (hence, obviously, $\check{\delta}^{0}\{f_{\alpha}\} \in \check{Z}^{1}(X, S^{1})$) we have that the first Chern class is one of the possible real lifts of $[\,\check{\delta}^{0}\{f_{\alpha}\}\,]_{S^{1}}$. Therefore, $p_{S^{1}} \, c_{1}(\{f_{\alpha}\})$ is the holonomy of the trivial line bundle on which the connection $A$, previously considered, is defined.

\subsubsection{Hypercohomological description}

The trivialized bundle $X \times \mathbb{C}$ with global connection $A$ corresponds to the hypercocycle $\{1, -A\} \in \check{Z}^{1}(X, \underline{S}^{1} \rightarrow \Omega^{1}_{\mathbb{R}})$. For $A$ flat and $\{f_{\alpha}\}$ parallel sections, we have $[\,\{1, -A\}\,] = [\,\{\check{\delta}^{0}f_{\alpha}, 0\}\,]$, thus the difference is a coboundary:
	\[\{1, -A\} \cdot \{\check{\delta}^{0}f_{\alpha}, \tilde{d}f_{\alpha}\} = \{\check{\delta}^{0}f_{\alpha}, 0\}
\]
thus $\tilde{d}f_{\alpha} = A_{\alpha}$ so that, as proven before, $[\,A\,] \simeq c_{1}(\{f_{\alpha}\})$.

If $f$ is globally defined, we get $\{1, -A\} \cdot \{1, \tilde{d}f\} = \{1, 0\}$ so that $[\,A\,] = [\,\tilde{d}f\,]$ which is integral: this corresponds to the choice of a global parallel section $f \cdot 1$ in $X \times \mathbb{C}$.

\subsection{Trivializations of gerbes}

Let us now consider a trivialization of a gerbe $\{h_{\alpha\beta}\} \in \check{C}^{1}(X, \underline{S}^{1})$ such that $\check{\delta}^{1}\{h_{\alpha\beta}\} \in \check{C}^{2}(X, S^{1})$. We can consider a connection $\{-A_{\alpha}\}$ such that $A_{\beta} - A_{\alpha} = \tilde{d}h_{\alpha\beta}$, as for an ordinary bundle. We have $dA_{\alpha} = dA_{\beta}$ so that $-F = -dA_{\alpha}$ is a global closed form whose de Rham class $[\,-F\,]$ is exactly the fractional Chern class of $[\,\{h_{\alpha\beta}\}\,] \in \check{\delta}^{-1}(\check{C}^{2}(X, S^{1})) \,/\, \check{B}^{1}(X, \underline{S}^{1})$. We define such a trivialization with connection as an element of the hypercohomology group:
	\[\check{H}^{1}\bigl(\, X, \underline{S}^{1} / S^{1} \overset{\tilde{d}}\longrightarrow \Omega^{1}_{\mathbb{R}} \,\bigr) \;.
\]
We interpret the Chern class of such trivializations as before: we consider the flat gerbe $[\,\{\check{\delta}^{1}h_{\alpha\beta}, 0, 0\}\,]$, and we represent it as $[\,\{1, 0, -F\}\,]$:
	\[\{1, 0, -F\} \cdot \{\check{\delta}^{1}h_{\alpha\beta}, -\tilde{d}h_{\alpha\beta} + A_{\beta} - A_{\alpha}, dA_{\alpha}\} = \{\check{\delta}^{1}h_{\alpha\beta}, 0, 0\}
\]
from which we obtain:
	\[A_{\beta} - A_{\alpha} = \tilde{d}h_{\alpha\beta} \qquad dA_{\alpha} = F\vert_{U_{\alpha}} \; .
\]
From these data we can now realize $F$ as a $\rm\check{C}$ech class: we have $F\vert_{U_{\alpha}} = dA_{\alpha}$ and $A_{\beta} - A_{\alpha} = \tilde{d}h_{\alpha\beta}$, thus $\check{\delta}^{1}\tilde{d}h_{\alpha\beta} = 0$, thus $(2\pi i)^{-1} \check{\delta}^{1}\log h_{\alpha\beta}$ is constant and expresses $[\,F\,]$ as $\rm\check{C}$ech class. The latter is exactly $c_{1}(\{h_{\alpha\beta}\})$.

\paragraph{}What happens for the holonomy of these connections? In general anyone of them is not well-defined as a function on closed curves, but it is a section of a line bundle that, on curves which are boundary of open surfaces, is canonically trivial and coincides with the one determined by the flat gerbe realized by $(1, 0, F)$ but with respect to the sections $\check{\delta}g$. In fact, the expression of the holonomy of $A$ on $\partial\Sigma$ coincides with the holonomy of $(\check{\delta}g, A_{\beta} - A_{\alpha}, dA_{\alpha})$ on $\Sigma$, but $\check{\delta}(g, 0) = (\check{\delta}g, d\log g_{\alpha\beta}, 0)$ and the sum is $(1, 0, dA_{\alpha})$, thus the gerbe is $(1, 0, F)$ but it is realized on open surfaces with respect to $\check{\delta}g$.

\section{Stack of coincident branes}

Up to now we have discussed the case of a single brane. In the case of a stack of coincident Dp-branes, we need non-abelian cohomology (see \cite{Brylinski}). However, here we would like to avoid a technical discussion and just state the main differences with respect to the abelian case. We will arrive to the same conclusions as \cite{Kapustin}, taking into account the presence of the Pfaffian.

\paragraph{}Let us consider again the fundamental equation \eqref{CoordChange}:
	\[\begin{split}
		\{g_{\alpha\beta\gamma}, &-\Lambda_{\alpha\beta}, B_{\alpha}\} \cdot \{g_{\alpha\beta\gamma}^{-1} \cdot \eta_{\alpha\beta\gamma}, \Lambda_{\alpha\beta}, dA_{\alpha}\} = \{\eta_{\alpha\beta\gamma}, 0, B+F\}\\
		&\{g_{\alpha\beta\gamma}^{-1} \cdot \eta_{\alpha\beta\gamma}, \Lambda_{\alpha\beta}, dA_{\alpha}\} = \{\check{\delta}^{1}h_{\alpha\beta}, -\tilde{d}h_{\alpha\beta} + A_{\beta} - A_{\alpha}, dA_{\alpha}\} \; .
	\end{split}
\]
Since $\check{\delta}^{1}h_{\alpha\beta} = g_{\alpha\beta\gamma}^{-1} \cdot \eta_{\alpha\beta\gamma}$, the class $[\,g^{-1}\eta\,] \in H^{1}(Y,\underline{S}^{1})$ must be trivial: this means that $\zeta\vert_{Y} = W_{3}(Y)$, which is the Freed-Witten anomaly equation. Instead, in the case of a stack of branes, $h_{\alpha\beta} \in U(n)$. Then, if we think of $g_{\alpha\beta\gamma}^{-1} \cdot \eta_{\alpha\beta\gamma}$ as a multiple of the identity $I_{n}$, the relation $\check{\delta}^{1}h_{\alpha\beta} = g_{\alpha\beta\gamma}^{-1} \cdot \eta_{\alpha\beta\gamma}$ is not a trivialization of $[\,g^{-1}\eta\,] \in H^{1}(Y,\underline{S}^{1})$ any more and it does not imply that $\zeta\vert_{Y} = W_{3}(Y)$. We thus rewrite the previous equation as:
\begin{equation}
\begin{split}
		&\{g_{\alpha\beta\gamma}, -\Lambda_{\alpha\beta}, B_{\alpha}\} \cdot \{g_{\alpha\beta\gamma}^{-1} \cdot \eta_{\alpha\beta\gamma}, \Lambda_{\alpha\beta}, d\tilde{A}_{\alpha}\} = \{\eta_{\alpha\beta\gamma}, 0, B+\tilde{F}\}\\
		&\{g_{\alpha\beta\gamma}^{-1} \cdot \eta_{\alpha\beta\gamma}, \Lambda_{\alpha\beta}, d\tilde{A}_{\alpha}\} = \textstyle\frac{1}{n} \Tr\, \{\check{\delta}^{1}h_{\alpha\beta}, -h_{\alpha\beta}^{-1}dh_{\alpha\beta} + h^{-1}_{\alpha\beta}A_{\beta}h_{\alpha\beta} - A_{\alpha}, dA_{\alpha} + A_{\alpha} \wedge A_{\alpha}\}
\end{split}
\end{equation}
where the trace is taken in all the components. We thus obtain $\tilde{A} = \frac{1}{n} \Tr A$ and $\tilde{F} = \frac{1}{n} \Tr F$.

\paragraph{}A rank-$n$ bundle $\{h_{\alpha\beta}\}$ such that $\check{\delta}^{1}\{h_{\alpha\beta}\}$ realizes a class in $H^{2}(X, \underline{S}^{1})$ is called a \emph{twisted bundle} or \emph{non-commutative bundle}. For $\beta$ the Bockstein homomorphism in degree 2 of the sequence $0 \rightarrow \mathbb{Z} \rightarrow \underline{\mathbb{R}} \rightarrow \underline{S}^{1} \rightarrow 0$, we define $\beta' = \beta[\,\check{\delta}^{1}\{h_{\alpha\beta}\}\,] \in H^{3}(X, \mathbb{Z})$. Thus, for the relation $\check{\delta}^{1}h_{\alpha\beta} = g_{\alpha\beta\gamma}^{-1} \cdot \eta_{\alpha\beta\gamma}$ to hold, one must have:
\begin{equation}
	\beta' = W_{3}(Y) - \zeta\vert_{Y} \; .
\end{equation}
This is the Freed-Witten anomaly equation for stack of branes. \emph{We remark that, while in the abelian case the $A$-field corresponds to a reparametrization of the gerbe, in the non-abelian case it provides another non-trivial gerbe}, which tensor-multiplies the gerbe of the B-field.

\paragraph{}The classification of configurations in this case is analogous to the case of a single brane, allowing for the possibility of a non-commutative bundle when $\beta' \neq 0$. For $\beta' = 0$, we have the same situation as before, with irrational Chern classes for non integral bundles.

\section{Conclusions}

We have classified the allowed configurations of $B$-field and $A$-field in type II superstring backgrounds with a fixed set of D-branes, which are free of Freed-Witten anomaly. For a single D-brane $Y \subset X$, we distinguish the following foundamental cases:
\begin{itemize}
	\item \emph{$B$ geometrically trivial, $w_{2}(Y) = 0$:} we fix the preferred gauge $(1, 0, 0)$, so that we have $(1, 0, F) = \delta(h, -A)$ with $(h, -A)$ a line bundle, up to the residual gauge symmetry;
	\item \emph{$B$ flat:} we fix the preferred gauge $(g,0,0)$ so that we have $(g^{-1}\eta,0,F) = \delta(h, -A)$ with $(h, -A)$ a ``bundle'' with, in general, a non-integral Chern class; the image in $S^{1}$ of such a Chern class is given by $\Hol(B\vert_{Y}) - w_{2}(Y)$; even if this bundle has integral Chern class, i.e., if $\Hol(B\vert_{Y}) = w_{2}(Y)$, in general it is defined only up to the torsion part; if $\Hol(B\vert_{Y}) = w_{2}(Y) = 0$ we end up with the previous case so that we recover the torsion part up to the residual gauge;
	\item \emph{$B$ generic:} we fix a gauge $(\eta, 0, B)$ so that we have $(1, 0, F) = \delta(h, -A)$ with $(h, -A)$ a \emph{non-canonical} line bundle, where non-canonicity is related to large gauge transformations $B \rightarrow B + \Phi$ and $F \rightarrow F - \Phi$ for $\Phi$ integral.
\end{itemize}
For a stack of coincident branes the situation is analogous, except for the possibility of non-commutative bundles.

\paragraph{}So far we have considered the case of one brane or stack of coincident branes. One may wonder what happens when we have more than one non-coincident branes or stacks of branes: this case is actually already included in the previous discussion, thinking of $Y$ as the disconnected union of all the world-volumes. In particular, the residual gauge symmetry becomes an ambiguity corresponding to the restriction to each brane of a \emph{unique} flat space-time bundle. In physical terms this can be seen as follows: if we choose two cycles, one for each brane, which are homologous in space-time but not necessarily homologically trivial, since the difference is homologically trivial we can link them by an open string loop stretching from one brane to the other. In this way we determine the holonomy on the difference, i.e., the difference of the holonomies on the two loops. We thus remain with a global uncertainty, represented by flat space-time line bundles.

\paragraph{}Let us briefly comment on the case of fractional branes coming from orbifolds. Using the notation of \cite{BDM}, let $\Gamma$ be the internal orbifold group, whose regular representation splits into $M$ irreducible representations of dimensions $d_{I}$ for $I = 0, \ldots, M-1$, and let $C_{I}$ be the corresponding cycles in the ADE-resolution of the orbifold singularity. $B$ is taken flat on the internal space and satisfying the formula $\int_{C_{I}} B = d_{I} \,/\, \abs{\Gamma}$ for $I = 1, \ldots, M-1$, while, on the last cycle, $\int_{C_{0}} B = -\sum_{I \neq 0} d_{I}\int_{C_{I}} B$. Moreover one chooses $F$ on a cycle representing $C_{0}$ (to be subsequently shrunk) such that $\int_{C_{0}}F = 1$, while, on the chosen representatives of the other cycles, one chooses $F = 0$. What does this mean in our language? One fixes a gauge $\{1, 0, B\}$ on the whole internal space ($\eta_{\alpha\beta\gamma} = 1$ because the manifold involved is spin), supposing that the hypercocycle fixed on the representatives of the $C_{I}$'s, for $I \neq 0$, is the restriction of the global one: one thus gets $F = 0$. On the representative of $C_{0}$, instead, we consider a hypercocycle corresponding to the restriction of the global one, modified by an automorphism of the gerbe which generates $F$ so that $\int_{C_{0}}F = 1$. In conclusion we obtain, on $C_{0}$, $\{1, 0, B + F\}$. This is \emph{not} the canonical gauge choice adopted in section \ref{GaugeTheories} that gives rise to a fractional bundle: had we made this choice, we would have obtained a bundle with a fractional Chern class $F$, whose imagine in $S^{1}$ is given by $\Hol(B\vert_{C_{I}}) = d_{I} \,/\, \abs{\Gamma}$.

\section*{Acknowledgements}

We would like to thank Jarah Evslin, Stefano Cremonesi and Francesco Benini for useful discussions.

\begin{flushleft} \huge
\vspace{1cm}
\textbf{Appendices}
\end{flushleft} \large

\appendix

\section{$\rm\bf\check{C}$ech Hypercohomology}\label{AppHyperC}

We refer to \cite{Brylinski} for a comprehensive treatment of hypercohomology. Given a sheaf $\mathcal{F}$ on a topological space $X$ with a good cover $\mathfrak{U} = \{U_{i}\}_{i \in I}$, we construct the complex of $\rm\check{C}$ech cochains:
	\[\check{C}^{0}(\mathfrak{U}, \mathcal{F}) \overset{\check{\delta}^{0}}\longrightarrow \check{C}^{1}(\mathfrak{U}, \mathcal{F}) \overset{\check{\delta}^{1}}\longrightarrow \check{C}^{2}(\mathfrak{U}, \mathcal{F}) \overset{\check{\delta}^{2}}\longrightarrow \cdots
\]
whose cohomology is by definition $\rm\check{C}$ech cohomology of $\mathcal{F}$. We recall, in particular, that $\check{\delta}^{p}: \check{C}^{p}(\mathfrak{U}, \mathcal{F}) \rightarrow \check{C}^{p+1}(\mathfrak{U}, \mathcal{F})$ is defined by $(\check{\delta}^{p}g)_{\alpha_{0} \cdots \alpha_{p+1}} = \sum_{i=0}^{p+1} (-1)^{i}g_{\alpha_{0} \cdots \check{\alpha}_{i} \cdots \alpha_{p+1}}$. If, instead of a single sheaf, we have a complex of sheaves:
	\[\cdots \overset{d^{i-2}}\longrightarrow \mathcal{F}^{i-1} \overset{d^{i-1}}\longrightarrow \mathcal{F}^{i} \overset{d^{i}}\longrightarrow \mathcal{F}^{i+1} \overset{d^{i+1}}\longrightarrow \cdots
\]
we can still associate to it a cohomology, called \emph{hypercohomology} of the complex. To define it, we consider the double complex made by the $\rm\check{C}$ech complexes of each sheaf:
\[\xymatrix{
	\vdots & \vdots & \vdots & \\
	\check{C}^{0}(\mathfrak{U}, \mathcal{F}^{q+1}) \ar[r]^{\check{\delta}^{0}} \ar[u]^{d^{q+1}} & \check{C}^{1}(\mathfrak{U}, \mathcal{F}^{q+1}) \ar[r]^{\check{\delta}^{1}} \ar[u]^{d^{q+1}} & \check{C}^{2}(\mathfrak{U}, \mathcal{F}^{q+1}) \ar[r]^{\phantom{XXX}\check{\delta}^{2}} \ar[u]^{d^{q+1}} & \cdots \\
	\check{C}^{0}(\mathfrak{U}, \mathcal{F}^{q}) \ar[r]^{\check{\delta}^{0}} \ar[u]^{d^{q}} & \check{C}^{1}(\mathfrak{U}, \mathcal{F}^{q}) \ar[r]^{\check{\delta}^{1}} \ar[u]^{d^{q}} & \check{C}^{2}(\mathfrak{U}, \mathcal{F}^{q})  \ar[r]^{\phantom{XXX}\check{\delta}^{2}} \ar[u]^{d^{q}} & \cdots \\
	\check{C}^{0}(\mathfrak{U}, \mathcal{F}^{q-1}) \ar[r]^{\check{\delta}^{0}} \ar[u]^{d^{q-1}} & \check{C}^{1}(\mathfrak{U}, \mathcal{F}^{q-1}) \ar[r]^{\check{\delta}^{1}} \ar[u]^{d^{q-1}} & \check{C}^{2}(\mathfrak{U}, \mathcal{F}^{q-1}) \ar[r]^{\phantom{XXX}\check{\delta}^{2}} \ar[u]^{d^{q-1}} & \cdots\\
	\vdots \ar[u]^{d^{q-2}} & \vdots \ar[u]^{d^{q-2}} & \vdots \ar[u]^{d^{q-2}} & 
	}
\]
We now consider the associated total complex\footnote{We use notation of \cite{Brylinski}, in which the two boundaries of the double complex commute, so that the boundary of the total complex has a factor $(-1)^{p}$. In the most common notation the two boundaries anticommute.}:
	\[T^{n} = \bigoplus_{p+q = n} \check{C}^{p}(\mathfrak{U}, \mathcal{F}^{q}) \qquad\quad d^{n} = \bigoplus_{p+q = n} \bigl(\,\check{\delta}^{p} + (-1)^{p}\,d^{q}\,\bigr)
\]
By definition, the \emph{$\rm\check{C}$ech hypercohomology} of the complex of sheaves is the cohomology of the total complex $H^{\bullet}(T^{n}, d^{n})$. It is denoted by:
	\[\check{H}^{\bullet}\bigl( \, \mathfrak{U}, \, \cdots \overset{d^{i-1}}\longrightarrow \mathcal{F}^{i} \overset{d^{i}}\longrightarrow \mathcal{F}^{i+1} \overset{d^{i+1}}\longrightarrow \cdots \, \bigr) \; .
\]

Using hypercohomology we can describe the group of line bundles with connection, up to isomorphism and pull-back of the connection, on a space $X$. We recall that a bundle with connection is specified by a couple $(\{h_{\alpha\beta}\}, \{A_{\alpha}\})$ where $\check{\delta}\{h_{\alpha\beta}\} = 1$ and $A_{\alpha} - A_{\beta} = (2\pi i)^{-1} d \log h_{\alpha\beta}$. The bundle is trivial if there exists a 0-cochain $\{f_{\alpha}\}$ such that $\check{\delta}^{0}\{f_{\alpha}\} = \{h_{\alpha\beta}\}$. Let us consider the complex of sheaves on $X$:
	\[\underline{S}^{1} \overset{\tilde{d}}\longrightarrow \Omega^{1}_{\mathbb{R}}
\]
where $\underline{S}^{1}$ is the sheaf of smooth $S^{1}$-valued functions, $\Omega^{1}_{\mathbb{R}}$ the sheaf of $1$-forms and $\tilde{d} = (2\pi i)^{-1} \, d\circ \log$. (The complex is trivially extended on left and right by $0$.) The associated $\rm\check{C}$ech double complex is given by:
\[\xymatrix{
	\check{C}^{0}(\mathfrak{U}, \Omega^{1}_{\mathbb{R}}) \ar[r]^{\check{\delta}^{0}} & \check{C}^{1}(\mathfrak{U}, \Omega^{1}_{\mathbb{R}}) \ar[r]^{\check{\delta}^{1}} & \check{C}^{2}(\mathfrak{U}, \Omega^{1}_{\mathbb{R}})  \ar[r]^{\phantom{XXX}\check{\delta}^{2}} & \cdots \\
	\check{C}^{0}(\mathfrak{U}, \underline{S}^{1}) \ar[r]^{\check{\delta}^{0}} \ar[u]^{\tilde{d}} & \check{C}^{1}(\mathfrak{U}, \underline{S}^{1}) \ar[r]^{\check{\delta}^{1}} \ar[u]^{\tilde{d}} & \check{C}^{2}(\mathfrak{U}, \underline{S}^{1}) \ar[r]^{\phantom{XXX}\check{\delta}^{2}} \ar[u]^{\tilde{d}} & \cdots
	}
\]
Thus we have that $\check{C}^{1}(\mathfrak{U}, \underline{S}^{1} \rightarrow \Omega^{1}_{\mathbb{R}}) = \check{C}^{1}(\mathfrak{U}, \underline{S}^{1}) \oplus \check{C}^{0}(\mathfrak{U}, \Omega^{1}_{\mathbb{R}})$. Given a line bundle $L \rightarrow X$ we fix a set of local sections, with respect to $\mathfrak{U}$, determining transition functions $\{g_{\alpha\beta}\}$ and local representation of the connection $\{A_{\alpha}\}$. We claim that $(g_{\alpha\beta}, -A_{\alpha}) \in \check{C}^{1}(\mathfrak{U}, \underline{S}^{1} \rightarrow \Omega^{1}_{\mathbb{R}})$ is a cocycle. In fact, by definition, $\check{\delta}^{1}(g_{\alpha\beta}, -A_{\alpha}) = (\check{\delta}^{1}g_{\alpha\beta}, -\tilde{d}g_{\alpha\beta} + \check{\delta}^{0}(-A_{\alpha}))$, thus cocycle condition gives $\check{\delta}^{1}g_{\alpha\beta} = 0$, i.e., $g_{\alpha\beta}$ must be transition functions of a line bundle, and $A_{\alpha} - A_{\beta} = (2\pi i)^{-1}d\log g_{\alpha\beta}$, the latter being exactly the gauge transformation of a connection. Moreover, coboundaries are of the form $\check{\delta}^{0}(g_{\alpha}) = (\check{\delta}^{0}g_{\alpha}, \tilde{d}g_{\alpha})$ and it is easy to prove that these are exactly the possible local representations of the trivial connection $\partial_{X}$ on the trivial bundle $X \times \mathbb{C}$, i.e., the unit element of the group of line bundles with connection. Thus, such group is isomorphic to:
	\[\check{H}^{1}(\mathfrak{U}, \underline{S}^{1} \overset{\tilde{d}}\longrightarrow \Omega^{1}_{\mathbb{R}}) \; .
\]

\section{Gerbes}\label{AppGerbes}

We refer to \cite{Hitchin} for a clear introduction to gerbes. A gerbe with connection is defined by a triple $(\{g_{\alpha\beta\gamma}\}, \{\Lambda_{\alpha\beta}\}, \{B_{\alpha}\})$ where $\check{\delta}\{g_{\alpha\beta\gamma}\} = 1$, $\check{\delta}^{1}\{\Lambda_{\alpha\beta}\} = \{(2\pi i)^{-1}d \log g_{\alpha\beta\gamma}\}$ and $B_{\alpha} - B_{\beta} = d\Lambda_{\alpha\beta}$. The gerbe is trivial if there exists a 1-cochain $\{f_{\alpha\beta}\}$ such that $\check{\delta}\{f_{\alpha\beta}\} = \{g_{\alpha\beta\gamma}\}$. We use the approach of \cite{Brylinski}. As the group of isomorphism classes of line bundles on $X$ is isomorphic to $\check{H}^{1}(X, \underline{S}^{1})$, the group of gerbes on $X$ up to isomorphism can be identified with $\check{H}^{2}(X, \underline{S}^{1})$. In this paper we consider this as the definition of gerbe.

We consider the complex of sheaves:
	\[\underline{S}^{1} \overset{\tilde{d}}\longrightarrow \Omega^{1}_{\mathbb{R}} \overset{d}\longrightarrow \Omega^{2}_{\mathbb{R}}
\]
where $\underline{S}^{1}$ is the sheaf of smooth $S^{1}$-valued functions, $\Omega^{p}_{\mathbb{R}}$ the sheaf of $p$-forms and $\tilde{d} = (2\pi i)^{-1} \, d\circ \log$. (The complex is trivially extended on left and right by $0$.) In analogy with the case of line bundles, we define the equivalence classes of gerbes with connection as the elements of the group:
	\[\check{H}^{2}(X, \underline{S}^{1} \rightarrow \Omega^{1}_{\mathbb{R}} \rightarrow \Omega^{2}_{\mathbb{R}}) \; .
\]
The $\rm\check{C}$ech double complex is given by:
\[\xymatrix{
	\check{C}^{0}(\mathfrak{U}, \Omega^{2}_{\mathbb{R}}) \ar[r]^{\check{\delta}^{0}} & \check{C}^{1}(\mathfrak{U}, \Omega^{2}_{\mathbb{R}}) \ar[r]^{\check{\delta}^{1}} & \check{C}^{2}(\mathfrak{U}, \Omega^{2}_{\mathbb{R}})  \ar[r]^{\phantom{XXX}\check{\delta}^{2}} & \cdots \\
	\check{C}^{0}(\mathfrak{U}, \Omega^{1}_{\mathbb{R}}) \ar[r]^{\check{\delta}^{0}} \ar[u]^{d} & \check{C}^{1}(\mathfrak{U}, \Omega^{1}_{\mathbb{R}}) \ar[r]^{\check{\delta}^{1}} \ar[u]^{d} & \check{C}^{2}(\mathfrak{U}, \Omega^{1}_{\mathbb{R}}) \ar[r]^{\phantom{XXX}\check{\delta}^{2}} \ar[u]^{d} & \cdots \\
	\check{C}^{0}(\mathfrak{U}, \underline{S}^{1}) \ar[r]^{\check{\delta}^{0}} \ar[u]^{\tilde{d}} & \check{C}^{1}(\mathfrak{U}, \underline{S}^{1}) \ar[r]^{\check{\delta}^{1}} \ar[u]^{\tilde{d}} & \check{C}^{2}(\mathfrak{U}, \underline{S}^{1}) \ar[r]^{\phantom{XXX}\check{\delta}^{2}} \ar[u]^{\tilde{d}} & \cdots
	}
\]
Thus we have that $\check{C}^{2}(\mathfrak{U}, \underline{S}^{1} \rightarrow \Omega^{1}_{\mathbb{R}} \rightarrow \Omega^{2}_{\mathbb{R}}) = \check{C}^{2}(\mathfrak{U}, \underline{S}^{1}) \oplus \check{C}^{1}(\mathfrak{U}, \Omega^{1}_{\mathbb{R}}) \oplus \check{C}^{0}(\mathfrak{U}, \Omega^{2}_{\mathbb{R}})$. By definition, $\check{\delta}^{1}(g_{\alpha\beta\gamma}, -\Lambda_{\alpha\beta}, B_{\alpha}) = (\check{\delta}^{2}g_{\alpha\beta\gamma}, \tilde{d}g_{\alpha\beta\gamma} + \check{\delta}^{1} (-\Lambda_{\alpha\beta}), -d(-\Lambda_{\alpha\beta}) + \check{\delta}^{0}B_{\alpha})$. Thus cocycle condition gives $\check{\delta}^{2}g_{\alpha\beta\gamma} = 0$, i.e., $g_{\alpha\beta\gamma}$ must be transition functions of a gerbe, and:
	\[\begin{split}
	&B_{\alpha} - B_{\beta} = d\Lambda_{\alpha\beta}\\
	&\Lambda_{\alpha\beta} + \Lambda_{\beta\gamma} + \Lambda_{\gamma\alpha} = (2\pi i)^{-1} d \log g_{\alpha\beta\gamma} \; .
\end{split}\]
Coboundaries are of the form $\check{\delta}^{1}(h_{\alpha\beta}, -A_{\alpha}) = (\check{\delta}^{1}h_{\alpha\beta}, -\tilde{d}h_{\alpha\beta} + \check{\delta}^{0}(-A_{\alpha}), d(-A_{\alpha}))$, thus gerbes of this form are geometrically trivial.

\end{document}